\documentclass[a4paper,11pt]{article}
\usepackage{jheppub} 
\usepackage[T1]{fontenc} 
\usepackage{graphicx}
\usepackage{tikz}
\usepackage{indentfirst}
\usepackage{slashed}
\usepackage{amsmath,amssymb,multirow,xspace,slashed,array,booktabs}
\usepackage{physics}
\usepackage{indentfirst}
\usepackage{url}
\usepackage{bm}
\usepackage{braket}
\allowdisplaybreaks[4]
\usepackage[colorlinks=true,urlcolor=blue,anchorcolor=blue,citecolor=blue,filecolor=blue,linkcolor=blue,menucolor=blue,pagecolor=blue]{hyperref}
\usepackage{natbib}
\setcitestyle{square, comma, numbers,sort&compress, super}
\usepackage{subfigure, placeins}
\usepackage{booktabs}
\usepackage{blindtext, rotating}
\usepackage{appendix}
\usepackage{afterpage}
\usepackage{enumitem}
\usepackage{marvosym}
\usepackage{bbold}
\usepackage{verbatim}
\usepackage{soul} 
\usepackage[normalem]{ulem}
\usepackage{pifont}
\usepackage{cancel}

\newcommand{\rmd}{\mathrm{d}}
\newcommand{\gev}{\mathrm{GeV}}
\newcommand{\tev}{\mathrm{TeV}}
\renewcommand{\vev}[1]{ \left\langle {#1} \right\rangle }

\title{\boldmath Axiogenesis from $SU(2)_R$ phase transition}

\author[a]{Keisuke Harigaya}
\author[b]{Isaac R. Wang}
\affiliation[a]{School of Natural Sciences, Institute for Advanced Study,\\
  Princeton, New Jersey, 08540 USA}
\affiliation[b]{New High Energy Theory Center,
  Department of Physics and Astronomy, Rutgers University, \\
  Piscataway, New Jersey, 08854-8019 USA}

\emailAdd{keisukeharigaya@ias.edu}
\emailAdd{isaac.wang@rutgers.edu}

\abstract{
The baryon asymmetry of the universe may be explained by rotations of the QCD axion in field space and baryon number violating processes. We consider the minimal extension of the Standard Model by a non-Abelian gauge interaction, $SU(2)_R$, whose sphaleron process violates baryon number. Assuming that axion dark matter is also created from the axion rotation by the kinetic misalignment mechanism, the mass scale of the $SU(2)_R$ gauge boson is fixed as a function of the QCD axion decay constant, and vise versa. Significant portion of the parameter space has already been excluded by new gauge boson searches, and the high-luminocity LHC will further probe the viable parameter space.
}

\begin{document}
\maketitle
\flushbottom

\newpage

\section{Introduction}

The Standard Model (SM) of particle physics has been established by the discovery of the SM-like Higgs and the precise measurements of its properties. Yet, our understanding on the weak and strong interactions seems incomplete. The weak CP violation from the quark Yukawa couplings would necessarily introduce an $\mathcal{O}(1)$ CP violating phase in the strong interaction~\cite{Bell:1969ts,Adler:1969gk,tHooft:1976rip}. However, the strong upper bound on the neutron electric dipole moment requires that the strong CP phase be smaller than $10^{-10}$~\cite{Crewther:1979pi,Baker:2006ts}. This apparent discrepancy is called the strong CP problem. 

The absence of the strong CP violation can be explained by a global $U(1)$ symmetry called the Peccei-Quinn (PQ) symmetry that has QCD anomaly and is spontaneously broken~\cite{Peccei:1977hh,Peccei:1977ur}.  
The angular direction of the complex field that spontaneously breaks the PQ symmetry is a Nambu-Goldstone boson and called the QCD axion~\cite{Weinberg:1977ma,Wilczek:1977pj}, which obtains a small mass by QCD strong dynamics. 

Due to its small mass, the QCD axion in general does not rest at the minimum of its potential in the early universe . Rather, it dynamically evolves and may play important cosmological roles. For example, oscillations of the axion field around the minimum of the potential may explain the dark matter of the universe~\cite{Preskill:1982cy,Abbott:1982af,Dine:1982ah}. 

The axion may rotate in field space.
The rotation is naturally induced by explicit breaking of the PQ symmetry by a higher dimensional interaction if the radial direction of the complex field takes on a large field value in the early universe~\cite{Co:2019wyp}, in analogy with the dynamics in Affleck-Dine baryogenesis~\cite{Affleck:1984fy}.
The kinetic energy of the rotation is eventually transferred into axion excitation around the minimum of the potential and becomes dark matter. This origin of axion dark matter is called the kinetic misalignment mechanism~\cite{Co:2019jts}.

The rotation of the axion field may also explain the baryon asymmetry of the universe, which is called axiogenesis~\cite{Co:2019wyp}.
The rotation corresponds to PQ charge asymmetry. The charge asymmetry is transferred into baryon asymmetry via strong and weak sphaleron processes. The baryon asymmetry is fixed after the electroweak phase transition because the weak sphaleron transition rate is exponentially suppressed.
In this minimal scenario, however, after requiring that axion dark matter not be overproduced by the kinetic misalignment mechanism, the amount of the baryon asymmetry produced by axiogenesis is smaller than the observed one.  

Further extensions of the SM may enhance the baryon asymmetry produced by axiogenesis. In such scenarios, the simultaneous explanation of dark matter and baryon asymmetry by the axion rotation connects the extensions of the SM with the QCD axion. For example, the baryon asymmetry may be enhanced if the electroweak phase transition occurs at a temperature higher than the prediction of the SM. In this scenario, the electroweak phase transition temperature, which would be correlated with masses of new particles that couple to the SM-like Higgs to modify the electroweak phase transition, is predicted as a function of the QCD axion decay constant~\cite{Co:2019wyp}.
Lepton number violation by a dimension-5 Majorana neutrino mass operator can also create baryon asymmetry~\cite{Domcke:2020kcp,Co:2020jtv} and is investigated in~\cite{Co:2020jtv} in the context of axion rotations. 

In this paper, we consider the minimal extension of the SM by a non-Abelian gauge interaction, $SU(2)_R$~\cite{Pati:1974yy,Mohapatra:1974gc}.
The resultant gauge symmetry is $SU(3)_c\times SU(2)_L\times SU(2)_R \times U(1)_X$.
The right-handed up and down quarks and the right-handed electrons and neutrinos are unified into $SU(2)_R$ doublets, and we may understand left- and right-handed fermions as well as weak gauge bosons in a more symmetric manner.
Also, the right-handed neutrinos may explain the SM neutrino mass by forming a Dirac partner with the left-handed ones.
(We will, however, also consider a setup where some of these motivations are obscured, such as the one where some of the right-handed charged leptons are embedded into $SU(2)_R$ singlets.)

The $SU(2)_R$ sphaleron process may also produce baryon asymmetry that is fixed after the $SU(2)_R$ phase transition. Naively, this does not seem to work, since the $SU(2)_R$ sphaleron process does not produce $B-L$ asymmetry. Then the baryon number produced by the $SU(2)_R$ sphaleron process is washed-out by the $SU(2)_L$ sphaleron process, so that the baryon asymmetry is fixed only after the electroweak phase transition. We present two scenarios where the wash-out is evaded. In one scenario, we introduce new chiral leptons, so that the $B-L$ symmetry has $SU(2)_R$ anomaly and is produced by the $SU(2)_R$ sphaleron process. In another scenario, we introduce new fermions with vector-like gauge charges. The total $B-L$ symmetry is still anomaly free, but if some of the new $B-L$ charged fermions decay only after the electroweak phase transition, the $B-L$ charge felt by the $SU(2)_L$ sphaleron is non-zero. This is in the same spirit as Dirac leptogenesis~\cite{Dick:1999je}. 
In both scenarios, after requiring that
the axion rotation explain the observed dark matter and baryon abundance, the mass of the new gauge boson is predicted as a function of the QCD axion decay constant, and vise versa, up to a model-independent $\mathcal{O}(1)$ constant.

This paper is organized as follows. In Sec.~\ref{sec:axiogenesis review}, we review axiogenesis and kinetic misalignment. In Sec.~\ref{sec:axiogenesis from su2}, we discuss how the $SU(2)_R$ phase transition can produce baryon asymmetry, evading the washout from the electroweak sphaleron process. We show the relation between the new gauge boson mass and the QCD axion decay constant. Sec.~\ref{sec:summary} provides a summary and discussion.

\section{Axiogenesis and kinetic misalignment}\label{sec:axiogenesis review}
In this section, we review the idea of the axion rotation in field space and how the rotation may produce baryon asymmetry. We then discuss the impact of the rotation on axion dark matter abundance.

\subsection{Axion rotation}\label{sec:axion rotation}
In field theoretical realization, the QCD axion $a$ arises from an angular direction of a complex PQ symmetry breaking field $P$,
\begin{align}
    P = \frac{1}{\sqrt{2}}S \times  {\rm exp}\left(i \frac{a}{S}\right),
\end{align}
where $S$ is a radial direction which we call the saxion. The field $P$ has a nearly $U(1)$ symmetric potential and obtains a non-zero field value $\vev{S}= f_a$. We will also use the angular variable $\theta\equiv a/S$.

It is usually assumed that the angular velocity of the axion field is negligible and does not affect the dynamics of the axion. This assumption may not be justified if the saxion takes on a large initial field value in the early universe. For large field values, higher dimensional terms of the potential are important, and some of them may explicitly break the $U(1)$ symmetry,
\begin{align}
    \Delta V = \frac{P^n}{M^{n-4}} + {\rm h.c.},
\end{align}
where $M$ is a mass scale.
The explicit breaking is likely, given that the PQ symmetry is anyway explicitly broken by the QCD anomaly and is at the best understood as an accidental symmetry~\cite{Holman:1992us,Barr:1992qq,Kamionkowski:1992mf,Dine:1992vx}, and that quantum gravity is expected to violate global symmetries~\cite{Giddings:1988cx,Coleman:1988tj,Gilbert:1989nq,Harlow:2018jwu,Harlow:2018tng}.
The higher dimensional term provides a potential gradient to the angular direction and drives angular motion. As the saxion field value decreases by the expansion of the universe, the higher dimensional terms become negligible and the $U(1)$ symmetry is approximately conserved. The field $P$ continues to rotate, preserving the angular momentum up to the dilution by the cosmic expansion. Such dynamics of a complex field was originally considered for scalar quarks and leptons in Affleck-Dine baryogenesis~\cite{Affleck:1984fy}. It is convenient to define the yield of the $U(1)$ charge,
\begin{align}
    Y_P \equiv \frac{\dot{\theta} S^2}{s},
\end{align}
where $s$ is the entropy density of the universe. Until the axion potential by the QCD strong dynamics that explicitly breaks the $U(1)$ symmetry becomes effective, $Y_P$ remains constant.

A large enough $Y_P$ for axiogenesis and kinetic misalignment requires that the potential of $S$ be flat, so that the rotation begins later and the charge density relative to the entropy density becomes larger. The flatness of the potential is natural in supersymmetric theories, where the potential of $S$ can vanish in the supersymmetric limit and be lifted by supersymmetry-breaking soft mass terms. See~\cite{Co:2019jts,Co:2019wyp} for details.

The motion of $P$ initiated by the above mechanism in general involves both angular and radial motion. As $P$ couples to the thermal bath, the radial motion eventually is dissipated. The angular motion, on the other hand, is not dissipated because of the approximate conservation of the $U(1)$ charge. One may wonder that the $U(1)$ charge can be transferred into particle-antiparticle asymmetry of excitation in the thermal bath, but it is free-energetically favored to keep almost all of the $U(1)$ charge in the form of the coherent rotation~\cite{Co:2019wyp}. After the completion of the thermalization, the ellipticity of the rotation becomes zero.

For the circular motion, the equation of motion of $P$ requires that the angular velocity of $P$ satisfy $\dot{\theta}^2 = V'(S)/S$. For $S\gg f_a$, this is as large as the curvature of the saxion potential, namely, the saxion mass. The evolution of $S$ and $\dot{\theta}$ can be derived from this relation and the conservation of the charge,
$\dot{\theta}S^2 \propto R^{-3}$,
 where $R$ is the scale factor of the universe. As $S$ decreases and eventually reaches near the minimal of its potential, $\dot{\theta}=\sqrt{V'(S)/S}$ begins to rapidly decrease in proportion to $R^{-3}$ and becomes much smaller than the saxion mass.

\subsection{Axiogenesis}\label{sec:axiogenesis}
The rotation of the axion produces baryon asymmetry~\cite{Co:2019wyp}. Although most of the $U(1)$ charge is stored in the form of the rotation, a small fraction of it is transferred into particle-antiparticle asymmetry of particle excitation in the thermal bath. The asymmetry may be transferred into baryon asymmetry through the electroweak sphaleron process. For the QCD axion, the PQ charge is transferred into the quark chiral asymmetry via the QCD anomaly and the strong sphaleron process, and the chiral asymmetry is transferred into the baryon asymmetry through the weak anomaly of the chiral asymmetry and the electroweak sphaleron process. Generic axion-like particles are discussed in~\cite{Co:2020xlh}. The production of the particle-antiparticle asymmetry can be also understood by an effective chemical potential provided by the non-zero velocity of the axion field treated as a background field~\cite{Domcke:2020kcp}, along the line of spontaneous baryogenesis~\cite{Cohen:1987vi,Cohen:1988kt}.

For a given temperature, the baryon asymmetry produced from the axion rotation, $n_{B,{\rm rot}}$, is given by
\begin{align}
    n_{B,{\rm rot}} = c_B \dot{\theta}T^2 = c_B \frac{T^2}{S^2} n_P,
\end{align}
where $c_B$ is a constant that depends on the detail of the model and is typically $\mathcal{O}(0.1)$. The baryon asymmetry normalized by the entropy density $s$ for a given temperature is
\begin{align}
\label{eq:B_axiogenesis}
Y_{B,{\rm rot}}\equiv \frac{n_{B,{\rm rot}}}{s} = \frac{45 c_B}{2\pi^2 g_*}  \frac{\dot{\theta}}{T} = c_B \frac{T^2}{S^2} Y_P,
\end{align}
where $g_*$ is the effective degree of freedom.
In the SM, the baryon asymmetry is frozen around $T\simeq 130$ GeV$\equiv T_{\rm sp,L}$~\cite{DOnofrio:2014rug} since the rate of the electroweak sphaleron process is exponentially suppressed after the electroweak phase transition. To obtain the final baryon asymmetry created by the axion rotation, we evaluate $Y_B$ in Eq.~(\ref{eq:B_axiogenesis}) at this temperature,
\begin{align}
\label{eq:YB}
    Y_{B,{\rm rot}} = & 9\times 10^{-11}\times \left(\frac{c_B}{0.1}\right) \left(\frac{\dot{\theta}(T=T_{\rm sp,L})}{5~{\rm keV}}\right) \frac{130~{\rm GeV}}{T_{\rm sp,L}} \nonumber \\
    =& 9\times 10^{-11} \times \left(\frac{c_B}{0.1}\right) \left(\frac{10^9~{\rm GeV}}{f_a}\right)^2 \left(\frac{T_{\rm sp,L}}{130~{\rm GeV}}\right)^2 \left(\frac{f_a}{S}\right)^2 \frac{Y_P}{5\times 10^4}.
\end{align}

\subsection{Kinetic misalignment mechanism}\label{sec:kinetic misalignment}

The rotation of the axion also affects axion abundance.
The conventional picture is that the axion field starts oscillations from a certain field value when the Hubble expansion rate becomes smaller than the axion mass, and the axion oscillation behaves as dark matter~\cite{Preskill:1982cy,Abbott:1982af,Dine:1982ah}. 
For sufficiently large angular momentum, the kinetic energy of the axion field is larger than the potential energy when the conventional oscillation would occur. The axion does not oscillate and continues to rotate in field space. The kinetic energy, if transferred into axions that eventually become non-relativistic,  may be the origin of the dark matter density of the universe~\cite{Co:2019jts}.

The original picture presented in~\cite{Co:2019jts} is that the rotation remains coherent and the axion field begin oscillations around the minimum of the potential when the kinetic energy becomes comparable to the potential energy. However, when the axion field moves in an anharmonic potential, parametric resonant production~\cite{Dolgov:1989us,Traschen:1990sw,Kofman:1994rk,Kofman:1997yn} of axion fluctuations may occur~\cite{Jaeckel:2016qjp,Berges:2019dgr,Fonseca:2019ypl}. For the cosine potential, the effective production rate is given by~\cite{Fonseca:2019ypl}
\begin{align}
    \Gamma \simeq \frac{m_a^4}{\dot{\theta}^3},
\end{align}
where $m_a$ is the axion mass that may depend on the temperature. As $\dot{\theta}$ decreases, the production rate becomes larger.
One can see that when the kinetic energy of the axion field would be larger than the potential energy when the conventional oscillation would occur ($\dot{\theta} > m_a \sim H $), the production rate becomes larger than the Hubble expansion rate before the kinetic energy becomes comparable to the potential energy ($\dot{\theta}\sim m_a$). Therefore, the rotation of the axion field ends by loosing its energy via the production of axion fluctuations, rather than simply by the kinetic energy becoming smaller than the potential energy via the cosmic expansion.

The produced axions are relativistic and have a momentum $\simeq$ energy $\simeq \dot{\theta}/2$. Dividing the energy density  $\dot{\theta}^2f_a^2/2$ by the energy of the produced axion quanta, we obtain the number density of the axions~\cite{Co:2021rhi},
\begin{align}
\label{eq:Ya}
   Y_a =  \frac{n_a}{s} = \frac{\rho_\theta }{s \dot{\theta}/2}  = \frac{\dot{\theta}f_a^2}{s} = Y_P.
\end{align}
The number density is similar to the one obtained in~\cite{Co:2019jts} based on the assumption of coherent motion. This is not by accident and is for a good reason; in both dynamics the axion energy per quanta at the time of production is around the natural energy scale of the rotation, $\dot{\theta}$.
When the parametric resonance becomes effective, number-changing axion self-scattering rates, including Bose enhancement, become as large as the Hubble expansion rate, but the scattering rates soon become smaller than the expansion rate because of the cosmic expansion and the strong dependence of the scattering rates on the axion number density. We thus expect that the reduction of the axion number density by self-scattering is at the most by $\mathcal{O}(1)$, and we adopt the estimation in Eq.~(\ref{eq:Ya}). The reduction factor can be precisely determined by performing lattice computation.

For the parameter region relevant for the QCD axion, the produced axions become cold enough to be dark matter of the universe by the cosmic expansion. The axion dark matter abundance is
\begin{align}
\frac{\rho_a}{s} = m_a Y_a = 0.4~{\rm eV} \frac{10^{9}~{\rm GeV}}{f_a} \frac{Y_P}{70},
\end{align}
while the observed dark matter abundance is $\rho_{\rm DM}/s \simeq 0.4$ eV.
Using Eq.~(\ref{eq:YB}), we obtain
\begin{align}
\label{eq:ratio}
\frac{\rho_{\rm a}}{\rho_{\rm DM}}= 700 \frac{f_a}{10^9~{\rm GeV}} \frac{0.1}{c_B} \left(\frac{130~{\rm GeV}}{T_{\rm sp,L}}\right)^2 \frac{Y_{B,{\rm rot}}}{Y_{B}}.
\end{align}
For the standard electroweak phase transition with $T_{\rm sp,L}=130$ GeV, the axion dark matter is overproduced even for the smallest $f_a$ satisfying the astrophysical lower bound, $f_a \sim 10^8$ GeV~\cite{Ellis:1987pk,Raffelt:1987yt,Turner:1987by,Mayle:1987as,Raffelt:2006cw,Chang:2018rso,Carenza:2019pxu}.

The overproduction may be avoided by beyond-the SM interactions that violate baryon or lepton numbers.
In the next section, we introduce the minimal extension of the SM gauge group by a non-Abalian gauge group $SU(2)_R$, whose sphaleron processes (effectively) violate $B-L$.

\section{Axiogenesis from $SU(2)_{R}$}
\label{sec:axiogenesis from su2}

In this section, we discuss how baryon asymmetry can be produced by axiogenesis with the aid of an $SU(2)_R$ phase transition.

\subsection{$SU(2)_R$ gauge symmetry and fermion masses}
\label{sec:SU2 gauge group}

\begin{table}[tbp]
\caption{The minimal fermion content of a theory with $SU(3)_c\times SU(2)_L\times SU(2)_R\times U(1)_X$ gauge symmetry.}
\begin{center}
\begin{tabular}{|c|c|c|c|c|} \hline
 & $q_i$ & $\bar{q}_i$ & $\ell_i$ & $\bar{\ell}_i$ \\ \hline
 $SU(3)_c$ & {\bf 3} & {\bf 3} & {\bf 1} & {\bf 1} \\
 $SU(2)_L$ & {\bf 2} & {\bf 1} & {\bf 2} & {\bf 1} \\
 $SU(2)_R$ & {\bf 1} & {\bf 2} & {\bf 1} & {\bf 2} \\
 $U(1)_X$ & $\frac{1}{6}$ & $-\frac{1}{6}$ & $- \frac{1}{2}$ & $\frac{1}{2}$ \\ \hline
\end{tabular}
\end{center}
\label{tab:fermions_min}
\end{table}%

The SM may be embedded into a theory with an extra non-Abelian gauge symmetry, $SU(2)_R$. The gauge group is $SU(3)_c\times SU(2)_L\times SU(2)_R\times U(1)_X$, and the $SU(2)_R\times U(1)_X$ part is broken down to $U(1)_Y$ at an energy scale above the electroweak scale. This can be done by, e.g., a scalar with a charge $({\bf 1},{\bf 1},{\bf 3}, -2)$ or that with $({\bf 1},{\bf 1},{\bf 2}, 1/2)$. We consider the latter and call it $H_R$. We consider the cases where the electroweak symmetry $SU(2)_L\times U(1)_Y$ is broken by $\Phi({\bf 1},{\bf 2},{\bf 2},0)$ or $H_L({\bf 1},{\bf 2},{\bf 1},-1/2)$. The standard fermion embedding is the SM quark and lepton $SU(2)_L$ doublets in $(q_i,\ell_i)$ and the SM $SU(2)_L$ singlets in $(\bar{q}_i,\bar{\ell}_i)$ whose gauge charges are shown in Table~\ref{tab:fermions_min}. We will also consider a setup where some of the SM right-handed fermions come from $SU(2)_R$ singlets.

For the electroweak symmetry breaking by $\Phi(\bm{1}, \bm{2}, \bm{2}, 0)$, the SM Yukawa couplings are given by
\begin{align}
\label{eq:SM bi-doublet yukawa}
\mathcal{L}_{\rm Yukawa} &=  y_{q}^{ij}\Phi q_i \bar{q}_i + \tilde{y}_{q}^{ij}\tilde{\Phi} q_i \bar{q}_i + y_{\ell}^{ij}\Phi \ell_i \bar{\ell}_i  + \tilde{y}_{\ell}^{ij}\tilde{\Phi} \ell_i \bar{\ell}_i + {\rm h.c.}.
\end{align}
This leads to the SM fermion masses
\begin{align}
\label{eq:bi-doublet fermion mass}
&m_{u}^{ij} = y_q^{ij}v_1 + \tilde{y}_{q}^{ij}v_2^*, \qquad m_{d}^{ij} = y_q^{ij}v_2 + \tilde{y}_{q}^{ij}v_1^*, \\[1ex]
    &m_{e}^{ij} = y_\ell^{ij}v_2 + \tilde{y}_{\ell}^{ij}v_1^*,
     \qquad m_{\nu}^{ij} = y_\ell^{ij}v_1 + \tilde{y}_{\ell}^{ij}v_2^*, \nonumber \\
& \langle \Phi \rangle =\begin{pmatrix} v_1 & 0 \\ 0 & v_2 \end{pmatrix},
\end{align}
where $\sqrt{v_1^2 + v_2^2} = 174 \gev$ is the SM Higgs vacuum expectation value (vev).

The SM neutrinos obtain Dirac masses with the right-handed neutrinos as large as the charged lepton masses unless fine-tuned.%
\footnote{The fine-tuning seems absent if $v_2 < 10^{-8} v_1$ and $|y_\ell^{ij}| < 10^{-10}$ (or $v_1 < 10^{-8} v_2$ and $|\tilde{y}_\ell^{ij}| < 10^{-10}$.) However, since we need both $y_q^{ij}$ and $\tilde{y}_q^{ij}$, quantum corrections from the quark Yukawa couplings necessarily generate non-zero $\Phi^2$ and $v_2$ cannot be that smaller than $v_1$.}
To avoid the fine-tuning, we may introduce singlet fields $S$ that obtain Dirac masses with the right-handed neutrinos via a Yukawa interaction $y H_R^{\dagger} S \bar{\ell}$. We find that this interaction does not change the amount of baryon asymmetry produced via axiogenesis by more than 1\%. The SM neutrino mass may be given by dimension-5 operators $(\Phi \ell)^2$ arising from UV physics, or by a small Majorana mass of $S$.

For the electroweak symmetry breaking by $H_L$, the SM Yukawa couplings may be given by dimension-5 operators,
\begin{align}
  \label{eq:SM quark yukawa}
\mathcal{L}_{{\rm Yukawa},q} =  &\,\, \frac{c_u^{ij}}{M_u}(H_L^\dag q_i) (H_R^\dag \bar{q}_j)
+ \frac{c_d^{ij}}{M_d}(H_L q_i) (H_R \bar{q}_j) + {\rm h.c.}, \\
\label{eq:SM lepton yukawa}
\mathcal{L}_{{\rm Yukawa},\ell} =  &\,\,\frac{c_e^{ij}}{M_e}(H_L \ell_i) (H_R \bar{\ell}_j) + {\rm h.c.},
\end{align}
with $y_{\rm SM} = c v_R/M$.
The dimension-5 operators may be obtained by exchange of heavy Dirac fermions $\Psi \bar{\Psi}$,
\begin{align}
\label{eq:UV completion}
    {\cal L} = x \psi H_L^{(\dag)}\bar{\Psi} + x' H_R^{(\dag)}\bar{\psi} \Psi + M_\Psi  \Psi \bar{\Psi} + {\rm h.c.},~~\frac{c}{M} = \frac{x x'}{M_\Psi},
\end{align}
where $\psi$ is $q$ or $\ell$.
For the top Yukawa coupling, the masses of the Dirac fermion must be around or below $v_R$.
In order for the dimension-5 description to be appropriate around the $SU(2)_R$ phase transition, $M\gtrsim v_R$ is required, and we mainly consider such a case.
One may instead take another limit of $M < x' v_R$, for which the SM fermion is dominantly $\bar{\Psi}$. We consider such a case in Sec.~\ref{subsec:extra singlet}.

The neutrino mass may be given by $H_L^\dag \ell H_R^\dag \bar{\ell}$. The right-handed neutrino masses are as large as the SM neutrino masses. The right-handed neutrinos are thermalized via $SU(2)_R\times U(1)_X$ gauge interactions and remain as dark radiation. For the experimentally allowed $v_R\gtrsim 10$ TeV, the decoupling of the right-handed neutrinos occurs before the QCD phase transition, and the amount of the right-handed neutrino dark radiation is $\Delta N_{\rm eff}\lesssim 0.3$. This amount of dark radiation is consistent with the current upper bound from the observations of the cosmic microwave background~\cite{Planck:2018vyg}, but will be probed by the next generation observations.
The amount of dark radiation may be reduced if entropy production occurs after the right-handed neutrinos decouple from the thermal bath. Such entropy production may come from the radial direction of the PQ symmetry breaking field.
We may instead introduce singlet fields $S$ and Yukawa interactions $y H_R^{\dagger} S \bar{\ell}$ to make the right-handed neutrinos heavy and decay via $W_R$ exchange.

\subsection{Baryon number violation by $SU(2)_R$ sphaleron}

With the minimal fermion content,
the baryon and lepton symmetry have mixed anomaly with $SU(2)_L$ and $SU(2)_R$,
\begin{align}
\partial J_{B,L} = -\frac{3}{32\pi^2}(g_L^2 W_{L\mu \nu}\tilde{W}_L^{\mu \nu} - g_R^2 W_{R \mu \nu}\tilde{W}_R^{\mu \nu}).
\end{align}
The $SU(2)_R$ sphaleron process is effective before the $SU(2)_R$ phase transition and may convert the particle asymmetry produced by the axion rotation into baryon asymmetry. One may wonder if we can replace $T_{\rm sp,L}$ in Eq.~(\ref{eq:ratio}) with the temperature at which the $SU(2)_R$ sphaleron process freezes-out, $T_{\rm sp,R}$.
An obvious obstacle to this idea is that the $SU(2)_R$ sphaleron process preserves the $B-L$ symmetry and only creates $B+L$ asymmetry. The $B+L$ asymmetry produced by the $SU(2)_R$ sphaleron process may be washed out by the $SU(2)_L$ sphaleron process. This is indeed the case for models with only the minimal fermion content in Table~\ref{tab:fermions_min}, where the Yukawa couplings rapidly transfer the $B+L$ asymmetry of $\bar{q}$ and $\bar{\ell}$ produced by the $SU(2)_R$ sphaleron process into that of $q$ and $\ell$ that is washed out by the $SU(2)_L$ sphaleron.%
\footnote{This may be avoided if the $SU(2)_R$ symmetry breaking scale is sufficiently high. The scattering by the electron Yukawa coupling is inefficient at temperatures above $10^{6}$ GeV. If the exchange of $W_R$ gauge boson becomes ineffective above this temperature, which is the case for $v_R\gtrsim 10^9$ GeV, the part of the $B+L$ asymmetry may be stored at the decoupled right-handed neutrinos in $\bar{\ell}$, so that other fermions effectively feel non-zero $B-L$ asymmetry. We do not consider this case in this paper.}
The resultant baryon asymmetry is still non-zero since the rotating axion continues to create particle asymmetry, but the baryon asymmetry is frozen at $T_{\rm sp,L}=130$ GeV, and axion dark matter is overproduced or baryon asymmetry is under-produced.

In this paper, we consider two classes of models that avoid the wash-out of the asymmetry created by the $SU(2)_R$ sphaleron process.
\begin{itemize}
    \item 
    In Sec.~\ref{sec:Chiral}, we consider a model with additional chiral fermions with lepton and $SU(2)_R$ charges. The $B-L$ symmetry then has a mixed anomaly $SU(2)_R$~\cite{Fujikura:2021abj} and the wash-out is avoided. 
    \item
    In Sec.~\ref{sec:EChiral}, we consider  models with additional fermions with vector-like gauge charges. The $B-L$ symmetry is still anomaly-free, but if some of $B$ or $L$ charged new particles decay only after the electroweak phase transition, the electroweak sphaleron effectively feels non-zero $B-L$ asymmetry and the wash-out is avoided.
\end{itemize}
In both models, non-zero (effective) $B-L$ asymmetry is frozen at $T=T_{\rm sp,R}$, and we obtain
\begin{align}
  \frac{n_B}{s} = C \left( \frac{n_{B-L,{\rm eff}}}{s}\right)_{T = T_{\rm sp,R}} \equiv \left(\frac{c_B\dot{\theta} T^2}{s}\right)_{T = T_{\rm sp,R}}.
\end{align}
Here $C=28/79\simeq 0.35$ if the right-handed neutrinos have masses above the electroweak scale. If they have masses below the electroweak scale, $W_R$ boson exchange maintains the chemical equilibrium of the right-handed neutrinos and $C=1/4=0.25$.
$c_B$ depends on the fermion contents and conservation laws but is typically $\mathcal{O}(0.1)$.
We show $c_B$ arising from the axion-gluon coupling for each model in Secs.~\ref{sec:Chiral} and \ref{sec:EChiral} for $C=1/4$. $c_B$ from the axion-$SU(2)_L$ or $SU(2)_R$ gauge boson coupling is shown in Appendix~\ref{sec:appendix of cB}.

The temperature $T=T_{\rm sp,R}$ is related with the mass of new gauge bosons that may be searched at collider experiments.
After requiring that axion dark matter be produced by the kinetic misalignment mechanism and the observed baryon asymmetry be produced by axiogenesis from $SU(2)_R$, we obtain a relation between the mass of the new gauge bosons and the QCD axion decay constant.
The model provides a novel connection between the QCD axion and collider searches for new particles. We provide this prediction in Sec.~\ref{sec:mass-fa} by estimating the relation between $T_{\rm sp,R}$ and the new gauge boson mass.

\subsection{Chiral matter}
\label{sec:Chiral}
In this subsection, we consider a model with anomalous $B-L$ under $SU(2)_R$.
This anomaly may be introduced by additional new chiral fermions with baryon and/or lepton charges. $B-L$ asymmetry is created from the axion rotation and $SU(2)_R$ sphaleron processes, and is frozen upon the $SU(2)_R$ phase transition. 
Specifically, we employ new fermions with a three-generation structure~\cite{Kim:2017qxo,Fujikura:2021abj}:
\begin{align}
\label{eq:EWBG model leptons}
&L_i =  \begin{pmatrix}
E_i \\ N_i
\end{pmatrix}
: ({\bf 1}, {\bf 1}, {\bf 2}, -\frac{1}{2})_{L=1} ,  \qquad \widetilde{L} = \begin{pmatrix}
\bar{\mathcal{E}} \\ \bar{\mathcal{X}}
\end{pmatrix}  : ({\bf {1}}, {\bf 1}, {\bf 2}, \frac{3}{2})_{L=-1} ,\nonumber \\[1ex]
&\bar{E}_i : ({\bf 1}, {\bf 1}, {\bf 1}, 1)_{L=-1} ,  \quad \bar{N}_i : ({\bf {1}}, {\bf 1}, {\bf 1}, 0)_{L=-1} ,\nonumber \\[1ex]
&\mathcal{E} : ({\bf 1}, {\bf 1}, {\bf 1}, -1)_{L=1} ,  \quad \mathcal{X} : ({\bf {1}}, {\bf 1}, {\bf 1}, -2)_{L=1},
\end{align}
where $i=1,2,3$ is the generation index.
One can see that $B-L$ has $SU(2)_R$ anomaly.
The masses of these fermions are given by Yukawa couplings with $H_R$,
\begin{align}
\label{eq:ewbg model yukawa}
    \mathcal{L}_R = y_E^{ij}H_R^{\dagger} L_i \bar{E}_j + y_N H_R L_i \bar{N}_j + y_{\mathcal{E}} H_R^{\dagger} \tilde{L} \mathcal{E} + y_{\mathcal{X}} H_R \tilde{L}\mathcal{X} + {\rm h.c.}
\end{align}

In order to obtain non-zero baryon asymmetry, the lepton number of the new fermions must be transferred into the SM lepton number.
This is achieved via portal couplings. If the SM Higgs is a doublet $H_L$, the possible portal couplings are
\begin{align}
  \label{eq:portal with HL}
  \mathcal{L}_{\rm int} =  g_{{E}}^{ij} H_L  \ell_i \bar{E}_j
+ g_N^{ij} H_L^\dagger \ell_i \bar{N}_j + g^i_{\bar{\mathcal{E}}} H_R \bar{\ell}_i  \, \mathcal{E} + m_L^{ij}L_i \bar{l}_j + \mathrm{h.c.}
\end{align}
If the SM Higgs comes from a bi-doublet $\Phi$, 
the first two terms involving $H_L$ are absent, and only the latter two may be present.

The portal couplings
must be effective before the electroweak phase transition, but does not have to be effective around the $SU(2)_R$ phase transition. If they are small and effective only after the $SU(2)_R$ phase transition, the $B-L$ number carried by the SM leptons is conserved around the $SU(2)_R$ phase transition. Non-zero $B-L$ totally comes from extra leptons,
\begin{align}
  \label{eq:baryon number in ewbg model}
  \frac{n_B}{s} = -\frac{1}{4} \frac{1}{s}\left(\sum_{i}\left(n_{L_i} - n_{\bar{E}_i} - n_{\bar{N}_i}\right) - n_{\tilde{L}} + n_{\mathcal{E}} + n_{\mathcal{X}}\right)_{T=T_{{\rm sp,R}}},
\end{align}
and we obtain $c_B \simeq 0.125$, as shown in Appendix~\ref{sec:appendix of cB}.
If the portal couplings are large and effective during $SU(2)_R$ phase transition, one should count the $B-L$ number carried by all the particles,
\begin{align}
    \label{eq: baryon number with portal coupling in ewbg model}
    \frac{n_B}{s} &= \frac{1}{4}\frac{1}{s}\left(\sum_i\left(\frac{n_{q_i} - n_{\bar{q}_i}}{3} - \left(n_{\ell_i} - n_{\bar{\ell_i}}\right)- \left(n_{L_i} - n_{\bar{E}_i} - n_{\bar{N}_i}\right)\right) + n_{\tilde{L}} - n_{\mathcal{E}} - n_{\mathcal{X}}\right)_{T=T_{{\rm sp,R}}}.
\end{align}
In this case, we find that $c_B$ ranges between $0.136$ and $0.354$; details are listed in Appendix~\ref{sec:appendix of cB}.
Relatively large $c_B\simeq 0.35$ is achieved because there are many fermions that can carry $B-L$ charges. When some of the portal couplings are turned off, there exist more conservation laws and the way the fermions carry asymmetry is restricted, so that $c_B$ goes down to $0.2$.

\subsection{Effectively chiral matter}
\label{sec:EChiral}

In this subsection, we consider models with additional fermions with vector-like gauge charges. The total $B-L$ is still anomaly-free, but the $B-L$ asymmetry of the SM fermions may be non-zero if the $B-L$ asymmetry of some of the extra fermions is transferred into the SM sector only after the electroweak phase transition.

\subsubsection{Right-handed SM fermions from singlets}
\label{subsec:extra singlet}
We first discuss a model with electroweak symmetry breaking by a doublet $H_L$. 
The SM Yukawa couplings~\eqref{eq:SM quark yukawa} and \eqref{eq:SM lepton yukawa} may be generated by the interactions in~\eqref{eq:UV completion}.
Since $\Psi \bar{\Psi}$ are vector-like fermions, the total $B-L$ is still anomaly-free.
Let us, however, assume that the Dirac mass $M_\Psi$ is small for some of the SM fermions.
Then the corresponding SM right-handed fermion is $\bar{\Psi}$ rather than $\bar{\psi}$. In this limit, after the $SU(2)_R$ sphaleron process freezes-out, the charges of $\psi \bar{\Psi}$ and $\bar{\psi} \Psi$ are separately conserved, and the $B-L$ asymmetry stored in $\bar{\psi} \Psi$ is not transferred into the SM sector. As a result, the SM fermion sector has non-zero $B-L$ asymmetry, and non-zero baryon asymmetry remains. $\bar{\psi} \Psi$ may decay after the electroweak phase transition.

We apply the above idea to the tau lepton.
The relevant fermions are
\begin{align}
  \label{eq: tau prime charge}
  \ell_3 (1, 2, 1, -\frac{1}{2})_{-1},\ \bar{\tau} (1, 1, 1, 1)_{-1},\ \tau^{\prime}(1, 1, 1, -1)_1,\ \bar{\ell}_3  (1,1,2, \frac{1}{2})_{-1}.
\end{align}
Here  $\ell_{3}$ is the SM left-handed third generation lepton doublet,
 $\bar{\tau}$ is the right-handed tau, and $\tau'$ and $\bar{\ell}_3$ are new particles.
The SM quarks and the first two generation of leptons are from the fermions in Table \ref{tab:fermions_min},
with the masses given by the dimension-5 terms in Eqs.~(\ref{eq:SM quark yukawa}) and (\ref{eq:SM lepton yukawa}) with the index $i, j = 1, 2, 3$ for quarks and $i, j = 1, 2$ for leptons.
The masses of the tau lepton and the new fermions are given by
\begin{align}
  \label{eq:tau mass}
  \mathcal{L}_{\tau} &= y_{\tau}H_L\ell_{3} \bar{\tau} + y_{\tau^{\prime}}(H_R\bar{\ell}_{3})\tau^{\prime} + {\rm h.c.}
\end{align}
The asymmetry of the new fermions can be transferred into SM particles via the following coupling and mass,
\begin{align}
  \label{eq:portal}
  \mathcal{L}_{\rm portal} = 
  g_{\tau}^{i}(H_{R}\bar{\ell}_{i})\tau^{\prime} + \epsilon \tau^{\prime}\bar{\tau},
\end{align}
where $i=1,2$.
As discussed at the beginning of this subsection, we require that these coupling and mass be sufficiently small so that the transfer occurs only after the electroweak phase transition. This may be guaranteed by an (approximate) $Z_2$ symmetry under which $\bar{\ell}_3$ and $\tau'$ are odd. Note that the model also works without the portal coupling nor mass; the asymmetry of the new particles is eventually kept in the neutral component of $\bar{\ell}_3$, but as long as the mass of it is much below the GeV scale, the asymmetric component does not lead to too large dark matter abundance.

In this model, $B-L$ carried by $\bar{\ell}_{3}$ and $\tau^{\prime}$
is non-zero and opposite to $B-L$ carried by other fermions, and the latter $B-L$ is responsible for the baryon asymmetry of the SM fermions. The baryon number is thus given by
\begin{align}
  \label{eq:baryon number in extra singlet model}
  \frac{n_B}{s} = -\frac{1}{4}\frac{1}{s}\left(n_{\bar{\ell}_3} - n_{\tau^{\prime}}\right)_{T=T_{{\rm sp,R}}}.
\end{align}
We find $c_B \simeq 0.147$, as shown in Appendix~\ref{sec:appendix of cB}.

The field content of the model is left-right symmetric, so we may readily embed the theory into a left-right symmetric theory. The hierarchy $v_L \ll v_R$ is achieved via soft breaking of the symmetry in the Higgs mass squared that may come from spontaneous breaking provided by a left-right odd field.%
\footnote{It is possible to achieve spontaneous symmetry breaking solely from radiative corrections to the potential of $H_L$ and $H_R$~\cite{Hall:2018let}, but that results in the formation of domain walls upon $SU(2)_R$ symmetry breaking and left-right symmetry breaking by another field at a high temperature is anyway necessary.}
The coupling $y_{\tau'}= y_\tau$, so the mass of the charged new particle $\tau'$ is predicted to be $m_\tau v_R/v_L$. At colliders, it is pair-produced by the hyper-charge gauge interaction and decays into the neutral component of $\bar{\ell}_3$, which is observed as missing energy, and a pair of SM fermions via $W_R$ exchange.

\subsubsection{Extra vector-like fermions}
\label{subsec:vectorlike model}

If the electroweak symmetry is broken by a bi-doublet $\Phi$, the structure in Eq.~(\ref{eq:UV completion}) is not applicable. We instead consider a structure that only requires $H_R$. (The model also works for electroweak symmetry breaking by $H_L$.) We can, for example, introduce vector-like lepton pairs $L$/$\bar{E}$ and $\bar{L}$/$E$ with the following charges,
\begin{align}
\label{eq:vectorlike new lepton}
L = (1,1,2,-\frac{1}{2})_{1}, \bar{E} = (1,1,1,1)_{-1}, \bar{L} = (1,1,2,\frac{1}{2})_{-1}, E = (1,1,1,-1)_{1}
\end{align}
The following Yukawa couplings of these leptons with $H_R$ give masses to them,
\begin{align}
\label{eq:vectorlike lepton yukawa}
\mathcal{L}_{vectorlike} = y_{L}L H_R^{\dagger} \bar{E} + y_{\bar{L}} \bar{L} H_{R} E.
\end{align}
We assume that the Dirac masses $m_L L \bar{L}$ and $m_E  E\bar{E}$ are small. 
The $(L, \bar{E})$ lepton may decay by $m_L^i L \bar{\ell}_i$, while  $(\bar{L}, E)$ may decay by $g_L^i \bar{L} \Phi \ell_i$ or $g_E^i \bar{\ell}_i H_R E$.

Since these two lepton pairs have opposite gauge charges and lepton numbers, the total $B-L$ is anomaly-free. 
However, non-zero SM baryon asymmetry may be produced if one of $(L,\bar{E})$ and $(\bar{L},E)$ decays via the portal coupling or mass only after the electroweak phase transition and the other pair does before the electroweak phase transition. (The model also works even if the former decay does not occur.) The latter can occur before or after the $SU(2)_R$ phase transition. If the latter decay is effective only after the $SU(2)_R$ phase transition, the baryon number is
\begin{align}
  \label{eq:baryon number in vectorlike model}
  \frac{n_B}{s} = \pm \frac{28}{79}\frac{1}{s}(n_L - n_{\bar{E}})|_{T=T_{{\rm sp,R}}} = \mp \frac{1}{4}\frac{1}{s}(n_{\bar{L}} - n_E)|_{T=T_{{\rm sp,R}}},
\end{align}
where the upper/lower sign is applicable for $(L,\bar{E})/(\bar{L},E)$ decaying after the electroweak phase transition.
We find $c_B \simeq 0.094$ as shown in Appendix~\ref{sec:appendix of cB}.
If $(\bar{L},E)$ decays via the portal coupling after the electroweak phase transition and the decay of $(L, \bar{E})$ by the portal mass is effective around the $SU(2)_R$ phase transition, the final baryon number is given by $n_B = -(1/4)(n_{\bar{L}} - n_E)$. If  $(L,\bar{E})$ decays after the electroweak phase transition and the decay of $(\bar{L},E)$ is effective around the $SU(2)_R$ phase transition, $n_B = (1/4)(n_L - n_{\bar{E}})$. Details are discussed in Appendix~\ref{sec:appendix of cB}. In either case, we find $c_B \simeq 0.151-0.161$.

\subsection{$SU(2)_{R}$ phase transition and sphaleron decoupling}
\label{subsec:phase transition}

As mentioned in sec~\ref{sec:SU2 gauge group}, the $SU(2)_R$ sphaleron process transfers the PQ charge into (effective) $B-L$ asymmetry, and the $B-L$ asymmetry is fixed after the $SU(2)_R$ sphaleron process decouples.
In this subsection, we discuss the $SU(2)_R$ phase transition and the sphaleron decoupling.

In the symmetric phase, the sphaleron transition rate per volume is given by~\cite{DOnofrio:2014rug}
\begin{align}
  \label{eq:sphaleron in symmetric phase}
  \Gamma_{\rm sph} \simeq 20 \alpha_R^5 T^4,
\end{align}
and the transition occurs rapidly. In the broken phase, on the other hand, the sphaleron rate is exponentially suppressed as~\cite{Fuyuto:2014yia,Funakubo:2009eg,Carson:1989rf,Carson:1990jm,Klinkhamer:1984di}
\begin{align}
    \Gamma_{\rm sph}(T) = \mathcal{A}(T) e^{-E_{\rm sph}/T},
\end{align}
where the sphaleron energy $E_{\rm sph} = 4\sqrt{2} \pi v_R B(\lambda_R/g_R^2)/g_R$. Here we use the convention where $H_R = (0, \phi)^t$ and $\phi = v_R$ at the minimum. The pre-factor $A(T)$ is given in Refs.~\cite{Fuyuto:2014yia,Funakubo:2009eg}. The function $B$ depends on $\lambda_R/g_R^2$, with $B(0) = 1.6$ and $B(\infty) = 2.7$. For $\lambda_R/g_R^2$ of our interest, it is around 2. The sphaleron process becomes ineffective when $\Gamma_{\rm sph} \leq H(T)$, where $H(T) \simeq 1.66 \sqrt{g_*}T^2/m_{\rm Pl}$ is the Hubble parameter, with $m_{\rm Pl} \simeq 1.22 \times 10^{19}~\gev$ the Planck mass. For example, around a reference point with $v_R = 15~\tev$, we obtain
\begin{align}
\label{eq:sphaleron decoupling condition}
    \frac{v_R(T_{\rm sp,R})}{T_{\rm sp,R}} \simeq 1.23 \left( \frac{2.0}{B} \right) g_R,
\end{align}
where $T_{\rm sp,R}$ is the decoupling temperature of the $SU(2)_R$ sphaleron process.

To obtain $T_{\rm sp,R}$ as a function of the model parameters, we calculate the effective potential via a perturbative method. We will comment on the validity of the perturbative computation later. Since $g_R v_R(T_{\rm sp,R}) \sim g_R^2 T_{\rm sp,R}$ is smaller than the temperature, we expect that the high-temperature expansion is a good approximation. (Indeed, we can reproduce the SM prediction on $T_{\rm sp,L}$ within a 10\% discrepancy in the high-temperature expansion.)
In this approximation, the general form of the effective potential $V_{\rm eff}$ of $H_R = (0, \phi_R)^{t}$ is~\cite{Quiros:1999jp,Morrissey:2012db}
\begin{align}
  \label{eq:Veff highT}
  V_{\rm eff} = D(T^2 - T_0^2)\phi_R^2 - E T \phi_R^3 + \lambda_R\phi_R^4.
\end{align}
We assume that the quartic coupling $\lambda_R$ is not small. Then the phase transition is not of the first order, so we may neglect the trilinear $E$ term. Furthermore, zero-temperature quantum corrections to the potential and finite-temperature corrections to $\lambda_R$ are negligible in comparison with the tree-level potential, so we may simply drop them.
We find that the resultant prediction on the $W_R$ boson mass coincides with that based on the full computation beyond the above approximations, shown in Appendix~\ref{sec:effective potential}, within 10\%.
The coefficients in Eq.~\eqref{eq:Veff highT} are given by
\begin{align}
  \label{eq:coefficients of Veff}
  D = \frac{1}{24 v_R^2}\left(6m_{W_{R}}^2 + 3 m_{Z^{\prime}}^2 + m_{H_R}^2 + \sum_f n_f m_{f}^2\right),~~
  T_0 = \sqrt{\frac{m_{H_R}^2}{2 D}},
\end{align}
where $m_{f}$ is the masses of fermions from the condensation of $H_R$. The analytical expression for $v_R(T)/T$ below the critical temperature is~\cite{Quiros:1999jp}
\begin{align}
  \label{eq:phi over T}
  \frac{v_R(T)}{T} =\sqrt{\frac{D}{2\lambda_R}(\frac{T_0^2}{T^2} - 1)}.
\end{align}
The relation between the vev of $H_R$ and the $SU(2)_R$ sphaleron decoupling temperature is
\begin{align}
    \label{eq:vR analytical}
    \frac{v_R(T=0)}{T_{\rm sp, R}} = \sqrt{\frac{D}{2\lambda_R} + (1.23 g_R)^2}.
\end{align}

The fermion contributions to $D$ depend on models. Numerically, unless there are several fermions with $\mathcal{O}(1)$ Yukawa couplings to $H_R$, the correction is dominated by the gauge boson contribution. For example, if there is a colored ($n_f=3$) fermion with a Yukawa coupling as large as the top quark Yukawa, the prediction on $T_{\rm sp,R}$ increases by about 40\%/20\% for $\lambda_R=0.1/0.3$.

In the above computation, we neglected the quartic coupling between $H_R$ and $H_L$ or $\Phi$. Because of their small degree of freedom, even if the quartic coupling is as large as unity, the prediction on $T_{\rm sp,R}$ changes only by 10\%.

We comment on the validity of the perturbative computation. When the sphaleron process is about to decouple, $\phi_R \simeq g_R T$ according to Eq.~(\ref{eq:sphaleron decoupling condition}). On the other hand, the expansion parameter for the corrections from the transverse components of the gauge bosons is $\sim g_R^2 T/(g_R \phi_R) \sim 1 $~\cite{Arnold:1992rz,Arnold:1994bp}, so the perturbative computation is only marginally valid. To estimate the uncertainty from the near-breakdown of the perturbativity, we vary the contribution of the transverse components of the gauge bosons to $D$ by a factor of two from the one-loop value. We find that $T_{\rm sp,R}$ varies by $10/5$\% for $\lambda_R = 0.1/0.3$, and this level of uncertainty is expected in the perturbative computation. The uncertainty is $\mathcal{O}(10)$\% despite the $\mathcal{O}(1)$ uncertainty in $D$ because $v_R(T_{\rm sp,R})$ is already close to $v_R(T=0)$.

\subsection{New gauge boson mass and axion decay constant}
\label{sec:mass-fa}

\begin{figure}[t]
  \centering
  \begin{minipage}[h]{0.49\linewidth}
    \includegraphics[width=2.5in]{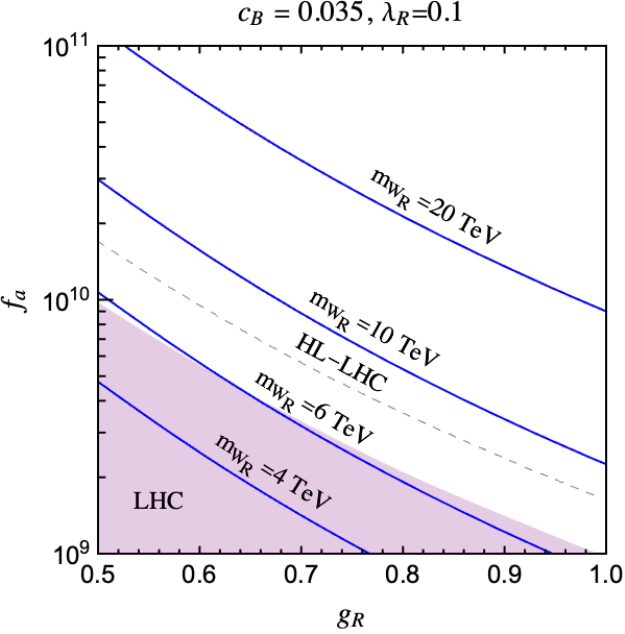}
  \end{minipage}
  \vspace{5mm}
   \begin{minipage}[h]{0.49\linewidth}
    \includegraphics[width=2.5in]{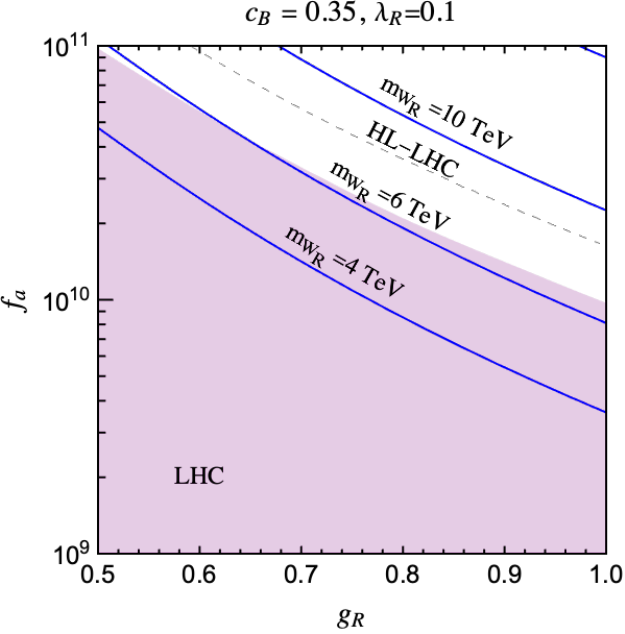}
  \end{minipage}
     \begin{minipage}[h]{0.49\linewidth}
    \includegraphics[width=2.5in]{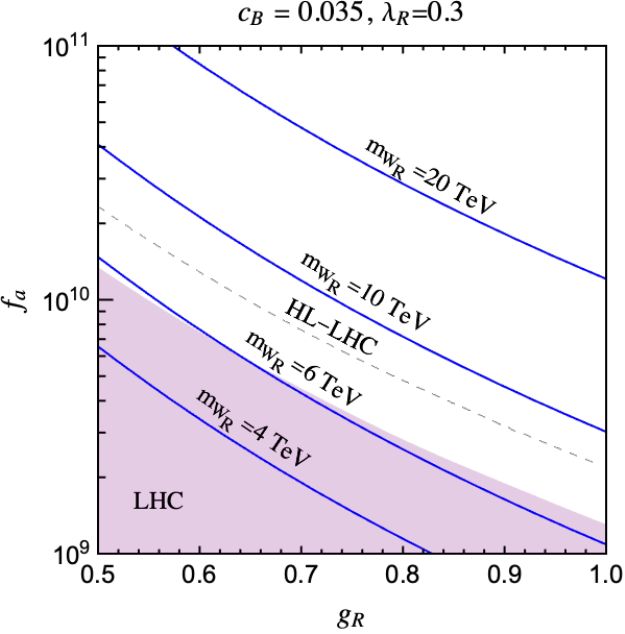}
  \end{minipage}
  \begin{minipage}[h]{0.49\linewidth}
    \includegraphics[width=2.5in]{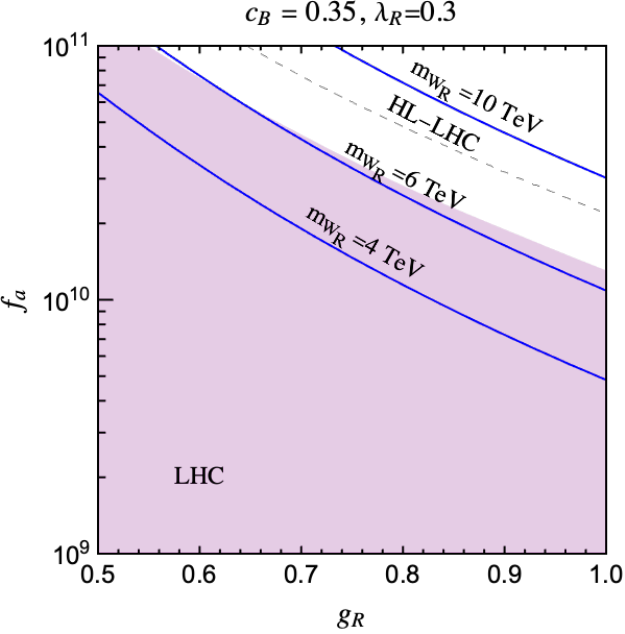}
  \end{minipage}
  \caption{Prediction on $m_{W_R}$ in the $(f_a,g_R)$ plane for $c_B=0.035$ and $0.35$. 
  The pink-shaded region is excluded by LHC search, while the gray dashed line shows the prospect of the HL-LHC.}
  \label{fig:W prediction}
\end{figure}

To explain both baryon and dark matter abundance from the axion rotation, it is required that
\begin{align}
  \label{eq:temperature requirement}
  T_{\rm sp,R} = (1.1~\tev) \left ( \frac{f_{a}}{10^{8}~\gev} \right )^{1/2} \left ( \frac{0.1}{c_{B}} \right )^{1/2}.
\end{align}
The $W_R$ boson mass is related with $f_{a}$,
\begin{align}
  \label{eq:prediction of WR}
  m_{W_R} = \frac{g_R v_R(T=0)}{\sqrt{2}} =(1.1~\tev) \left ( \frac{g_{R}}{\sqrt{2}} \right ) \left ( \frac{v_{R}(T=0)}{T_{\rm sp,R}} \right ) \left ( \frac{f_{a}}{10^{8}~\gev} \right )^{1/2} \left ( \frac{0.1}{c_{B}} \right )^{1/2},
\end{align}
with $v_R(T=0)/T_{\rm sp,R}$ analytically given by Eq.~(\ref{eq:vR analytical}) in the high-temperature expansion approximation, or given numerically without using the approximation. As mentioned above, the prediction based on the approximation is in good agreement with the numerical result using the full form of the thermal potential within 10\%. The contours of the $W_R$ boson mass in the $(f_a,g_R)$ plane for
$\lambda_R=0.1$ and $0.3$ and $c_B=0.035$ and $0.35$ together with LHC constraints~\cite{Aad:2019wvl} and HL-LHC prospects~\cite{ATLAS:2018tvr} are shown in Fig~\ref{fig:W prediction}.
$c_B$ around $0.035$ is a typical value when the PQ symmetry has QCD, $SU(2)_L$, and $SU(2)_R$ anomaly with the same anomaly coefficient and the number of extra particles is not large, but $c_B$ around $0.35$ (or even $1$) is also possible in the chiral model with a large number of extra particles and without extra conservation laws.
For $f_a \gtrsim 10^{11}$ GeV, the contribution to axion dark matter abundance from the conventional misalignment mechanism~\cite{Preskill:1982cy,Abbott:1982af,Dine:1982ah} dominates~\cite{Co:2019jts} and the assumption of axion dark matter from the kinetic misalignment mechanism breaks down.
If the HL-LHC finds a new gauge boson, that will point towards the QCD axion decay constant of $10^{9}\mathchar`-10^{11}$ GeV for $c_B=0.035\mathchar`-0.35$.

So far we assume that the QCD axion dark matter dominantly comes from the kinetic misalignment mechanism. When the rotation of the PQ symmetry breaking field is induced by a large field value of the radial direction, the initial rotation, before being thermalized, is not completely circular, and axions may be produced by parametric resonance~\cite{Co:2017mop,Co:2020dya,Co:2020jtv}; see~\cite{Co:2020dya,Co:2020jtv} for the estimation of the production rate of axions from a rotating PQ symmetry breaking field. If these axions are not thermalized subsequently, they also contribute to dark matter. In this case, the prediction on $m_{W_R}$ as a function of $f_a$ shown in Fig.~\ref{fig:W prediction} is understood as a lower bound, while $f_a$ as a function of $m_{W_R}$ is understood as an upper bound.

\section{Summary and discussion}\label{sec:summary}

The rotation of the QCD axion in field space may create both dark matter and baryon asymmetry of the universe.
Axion dark matter is created by the kinetic misalignment mechanism. The rotation of the axion corresponds to the Peccei-Quinn charge asymmetry, which is transferred into baryon asymmetry by the strong sphaleron process and baryon number violating processes. In the minimal scenario with the baryon number violation by the weak sphaleron, the baryon number is frozen at the $SU(2)_L$ phase transition temperature. For the standard $SU(2)_L$ phase transition temperature, after fixing the axion dark matter abundance, the amount of the baryon asymmetry created from the axion rotation is smaller than the observed one.

We considered the minimal extension of the SM gauge group by a non-Abelian gauge group $SU(2)_R$. The baryon number is also violated by the $SU(2)_R$ sphaleron process.
With the minimal fermion contents, $B-L$ is anomaly-free and the baryon number is still fixed at the $SU(2)_L$ phase transition. With additional fermions, however, the $B-L$ symmetry (effectively) has $SU(2)_R$ anomaly and the baryon asymmetry may be frozen at the $SU(2)_R$ phase transition temperature. 
Apart from model-dependent $\mathcal{O}(1)$ factors, the theory has three basic free parameters: the axion decay constant $f_a$, the new gauge boson mass $m_{W_R}$, and the angular velocity of the axion $\dot{\theta}$. After fixing dark matter and baryon abundances, there is only one free-parameter. The new gauge boson mass scale is predicted as a function of the axion decay constant, or vise versa, as is shown in Fig.~\ref{fig:W prediction}.

In this paper, we assume that the $SU(2)_R$ sphaleron is the dominant process that produces (effective) $B-L$ asymmetry.
If the generation of neutrino masses involves lepton number violation, it may also contribute to the production of $B-L$ asymmetry~\cite{Domcke:2020kcp,Co:2020jtv}.
In the models with $H_L$, the neutrino mass may be given by a lepton number-conserving Dirac mass $H_L^\dag \ell H_R^\dag \bar{\ell}$ and does not produce $B-L$. In the models with $\Phi$, to avoid too large Dirac masses from the Yukawa coupling $\ell \Phi \bar{\ell}$ or $\ell \tilde{\Phi} \bar{\ell}$, we introduce singlets $S$ that obtain Dirac masses with right-handed neutrinos via $S \bar{\ell}H_R^\dag$. The SM neutrino mass may be given by dimension-5 terms $(\Phi\ell)^2$. 
The effect of the possible dimension-5 Majorana neutrino mass terms is investigated in~\cite{Co:2020jtv}, and it is found that if the reheating temperature of the universe is sufficiently high and the saxion mass is above $10-1000$ TeV, the observed amount of baryon asymmetry may be explained by the lepton number violation from the dimension-5 interactions. In order for our prediction on $(m_{W_R},f_a)$ to hold in the presence of the dimension-5 operator, these conditions must be violated or the asymmetry produced by the dimension-5 operator must be washed out. The latter in fact occurs for the chiral model presented in Sec.~\ref{sec:Chiral} with portal couplings, since only the gauge charges are conserved around the $SU(2)_R$ phase transition. It is also possible to introduce small Majorana masses of $S$, which generate Majorana SM neutrino masses. In this case, the lepton number violation by the Majorana masses of $S$ can also produce $B-L$ asymmetry. However, since the required Majorana masses to reproduce the observed SM neutrino masses are small, the lepton number violation is not in thermal equilibrium and does not affect our scenario.

We discussed axiogenesis by $SU(2)_R$ phase transition. The mechanism can be readily generalized to phase transition in more generic non-Abelian gauge theories. Of course, to have observable collider signals, the new gauge boson should couple to some of the SM particles. The $SU(2)_R$ gauge symmetry would be the minimal example.

\section*{Acknowledgements}
R.W.~is grateful to Claudius Krause for useful discussions.
This work was supported in part by DOE grant DOE-SC0010008 (R.W.) and Friends of the Institute for Advanced Study (K.H.).

\appendix

\section{Calculation of $c_B$}
\label{sec:appendix of cB}

In this appendix, we calculate the coefficient $c_B$ by computing the asymmetry of each particle at thermal equilibrium. We work in the limit where the up and down Yukawa couplings are much smaller than other quark Yukawa couplings. In this limit, a linear combination of the PQ symmetry and the first generation quark chiral asymmetry remains unbroken. We may then obtain the amount of particle asymmetry at thermal equilibrium simply by requiring that each term in the transport equation vanish while imposing appropriate conservation laws.

The $\dot{\theta}$ dependence in the transport equations are derived following the method in~\cite{Co:2019wyp}: we consider the charge transfer between $\dot{\theta}f_a^2$ and fermions, and obtain the free-energy as a function of the transferred charges. 
Then $\dot{\theta}$-dependent terms are written down by following the principle of detailed balance.

\subsection{Extra chiral fermions}

For the model with extra chiral fermions described in Sec.~\ref{sec:Chiral}, we consider two cases where the SM Higgs is $H_L$ or from $\Phi$.

\subsubsection*{SM Higgs from $H_L$}
The Yukawa couplings of the SM fermions are given by dimension-5 operators such as $ (y/v_R) q\bar{q} H_L H_R$, giving a scattering rate $\sim y^2 T^3 / v_{R}^2$. Since we are interested in the transport equation around the $SU(2)_R$ phase transition, we take $v_R = T$.
The transport equations are then 
\begin{align}
  \label{eq:boltzmann eq for EWBG model with HL}
  \dot{n}_{q_i} &= \alpha_3\sum_{j}\left (y_{u}^{ij}\right )^{2}T \left (-\frac{n_{q_{i}}}{6} -  \frac{n_{\bar{q}_{j}}}{6} + \frac{n_{H_{L}}}{4} + \frac{n_{H_R}}{4}\right ) \\
  &+ \alpha_3\sum_{j}\left (y_{d}^{ij}\right )^{2} T \left (-\frac{n_{q_{i}}}{6} -  \frac{n_{\bar{q}_{j}}}{6} - \frac{n_{H_{L}}}{4} - \frac{n_{H_R}}{4}\right )\nonumber \\
  &+2 \Gamma_{\rm ss}\left (\sum_{j}\left (- n_{q_j} - n_{\bar{q}_j}\right ) - \frac{1}{2}\dot{\theta}T^2\right ) + 3 \Gamma_{\rm ws}\left (\sum_{j}\left (-n_{q_j} - n_{\ell_j}\right ) - \frac{c_{L}}{3}\dot{\theta}T^{2}\right) , \nonumber \\
  \dot{n}_{\bar{q}_i} &= \alpha_3\sum_{j}\left (y_{u}^{ji}\right )^{2}T\left (-\frac{n_{q_{j}}}{6} -  \frac{n_{\bar{q}_{i}}}{6} + \frac{n_{H_{L}}}{4} + \frac{n_{H_R}}{4}\right ) \nonumber \\
  & +\alpha_3\sum_{j} \left(y_{d}^{ji}\right )^{2}T\left (-\frac{n_{q_{j}}}{6} -  \frac{n_{\bar{q}_{i}}}{6} - \frac{n_{H_{L}}}{4} - \frac{n_{H_R}}{4}\right )\nonumber \\
  &+2 \Gamma_{\rm ss}\left (\sum_{j}\left (- n_{q_j} - n_{\bar{q}_j}\right ) - \frac{1}{2}\dot{\theta}T^2\right) + 3 \Gamma_{\rm rs}\left(\sum_j \left (- n_{\bar{q}_j} - n_{\bar{\ell}_j} - n_{L_j}\right ) - n_{\tilde{L}} - \frac{c_{R}}{3} \dot{\theta} T^2\right), \nonumber \\
  \dot{n}_{\ell_i} &=\left (\alpha_L + \alpha_R \right ) \left(y_e^{ii}\right )^2 T \left (-\frac{n_{\ell_i}}{2} - \frac{n_{\bar{\ell}_i}}{2} - \frac{n_{H_L}}{4} - \frac{n_{H_R}}{4}\right ) + \Gamma_{\rm ws} \left(\sum_j \left (-n_{q_j} - n_{\ell_j}\right ) - \frac{c_L}{3} \theta T^2\right)\nonumber  \\ 
  &+ \alpha_L \sum_j \left (g_E^{ij}\right )^2 T \left(- \frac{n_{\ell_i}}{2} - n_{\bar{E}_j} - \frac{n_{H_L}}{4}\right ) + \alpha_L \sum_j \left(g_N^{ij}\right )^2 T \left(- \frac{n_{\ell_i}}{2} - n_{\bar{N}_j} + \frac{n_{H_L}}{4}\right ), \nonumber\\
  \dot{n}_{\bar{\ell}_i} &= \left(\alpha_L + \alpha_R\right)\left (y_e^{ii}\right)^2 T \left(-\frac{n_{\ell_i}}{2} - \frac{n_{\bar{\ell}_i}}{2} - \frac{n_{H_L}}{4} - \frac{n_{H_R}}{4}\right)\nonumber \\
  &+ \Gamma_{\rm rs}\left(\sum_j \left (- n_{\bar{q}_j} - n_{\bar{\ell}_j} - n_{L_j}\right) - n_{\tilde{L}} - \frac{c_{R}}{3} \dot{\theta} T^2\right )\nonumber \\
  &+ \alpha_R g_{\mathcal{E}}^2 T \left( -\frac{n_{\bar{\ell}_i}}{2} - n_{\mathcal{E}} - \frac{n_{H_R}}{4}\right ) + \alpha_R \sum_j \frac{(m_L^{ij})^2}{T} \left(-\frac{n_{\bar{\ell}_i}}{2} - \frac{n_{L_j}}{2}\right ),\nonumber \\
  \dot{n}_{L_i} &= \alpha_R \sum_j \left(y_E^{ij}\right )^2 T \left (-\frac{n_{L_i}}{2} - n_{E_j} + \frac{n_{H_R}}{4}\right ) + \alpha_R \sum_j \left (y_N^{ij}\right )^2 T \left (-\frac{n_{L_i}}{2} - n_{N_j} - \frac{n_{H_R}}{4}\right )\nonumber \\
  & + \Gamma_{\rm rs}\left (\sum_j \left (- n_{\bar{q}_j} - n_{\bar{\ell}_j} - n_{L_j}\right ) - n_{\tilde{L}} - \frac{c_{R}}{3} \dot{\theta} T^2\right )\nonumber\\
  & + \alpha_R \sum_j \frac{(m_L^{ij})^2}{T}\left (-\frac{n_{\bar{\ell}_j}}{2} - \frac{n_{L_i}}{2}\right),\nonumber \\
  \dot{n}_{\bar{E}_i} &= \alpha_R \sum_j \left (y_E^{ji}\right )^2 T \left (-\frac{n_{L_j}}{2} - n_{\bar{E}_i} + \frac{n_{H_R}}{4}\right )
   + \alpha_L \sum_j \left (g_E^{ji}\right )^2 T \left (-\frac{n_{\ell_j}}{2} - n_{\bar{E}_i} - \frac{n_{H_L}}{4}\right ),\nonumber \\
  \dot{n}_{\bar{N}_i} &= \alpha_R \sum_j \left (y_N^{ji}\right )^2 T \left (-\frac{n_{L_j}}{2} - n_{\bar{N}_i} - \frac{n_{H_R}}{4}\right ) + \alpha_L \sum_j \left (g_N^{ji}\right )^2 T \left (-\frac{n_{\ell_j}}{2} - n_{\bar{N}_i} + \frac{n_{H_L}}{4}\right ), \nonumber\\
  \dot{n}_{\tilde{L}} &= \alpha_R \left (y_{\mathcal{E}}\right )^2 T \left (-\frac{n_{\tilde{L}}}{2} - n_\mathcal{E} + \frac{n_{H_R}}{4}\right )+ \alpha_R \left (y_{\mathcal{X}}\right )^2 T\left (-\frac{n_{\tilde{L}}}{2} - n_\mathcal{X} - \frac{n_{H_R}}{4}\right )\nonumber \\
  &+ \Gamma_{\rm rs}\left (\sum_j \left (- n_{\bar{q}_j} - n_{\bar{\ell}_j} - n_{L_j}\right ) - n_{\tilde{L}} - \frac{c_{R}}{3} \dot{\theta} T^2\right ), \nonumber\\
  \dot{n}_{\mathcal{E}} &= \alpha_R \left (y_{\mathcal{E}}\right)^2 T \left (-\frac{n_{\tilde{L}}}{2} - n_\mathcal{E} + \frac{n_{H_R}}{4}\right ) + \alpha_L \left (g_{\bar{\mathcal{E}}}^i\right )^2 T \left (- \frac{n_{\bar{\ell}_i}}{2} - n_\mathcal{E} - \frac{n_{H_R}}{4}\right ), \nonumber \\
  \dot{n}_{\mathcal{X}} &= \alpha_R \left (y_{\mathcal{X}}\right )^2 T \left (-\frac{n_{\tilde{L}}}{2} - n_\mathcal{X} - \frac{n_{H_R}}{4}\right ), \nonumber\\
  \dot{n}_{H_L} &= -\alpha_3\sum_{i,j}\left (y_{u}^{ij}\right )^{2}T \left (-\frac{n_{q_{i}}}{6} -  \frac{n_{\bar{q}_{j}}}{6} + \frac{n_{H_{L}}}{4} + \frac{n_{H_R}}{4} \right ) \nonumber \\
  &+ \alpha_3\sum_{i,j}\left (y_{d}^{ij} \right )^{2}T \left (-\frac{n_{q_{i}}}{6} -  \frac{n_{\bar{q}_{j}}}{6} - \frac{n_{H_{L}}}{4} - \frac{n_{H_R}}{4} \right )\nonumber \\
  & + \left (\alpha_L + \alpha_R\right )\sum_{i} \left (y_e^{ii}\right )^2 T \left (-\frac{n_{\ell_i}}{2} - \frac{n_{\bar{\ell}_i}}{2} - \frac{n_{H_L}}{4} - \frac{n_{H_R}}{4}\right )\nonumber \\
  & + \alpha_L \sum_{ij} \left (g_E^{ij}\right )^2 T \left (-\frac{n_{\ell_i}}{2} - n_{\bar{E}_j} - \frac{n_{H_L}}{4}\right )  + \alpha_L \sum_{ij} \left (g_N^{ij}\right )^2 T \left (-\frac{n_{\ell_i}}{2} - n_{\bar{N}_j} + \frac{n_{H_L}}{4}\right ), \nonumber \\
  \dot{n}_{H_R} & = -\alpha_3\sum_{ij}\left (y_{u}^{ij}\right )^{2}T \left (-\frac{n_{q_{i}}}{6} -  \frac{n_{\bar{q}_{j}}}{6} + \frac{n_{H_{L}}}{4} + \frac{n_{H_R}}{4}\right ) \nonumber \\
  &+ \alpha_3\sum_{ij}\left (y_{d}^{ij}\right )^{2}T\left (-\frac{n_{q_{i}}}{6} -  \frac{n_{\bar{q}_{j}}}{6} - \frac{n_{H_{L}}}{4} - \frac{n_{H_R}}{4}\right )\nonumber \\
  & + \left (\alpha_L + \alpha_R\right )\sum_{i} \left (y_e^{ii}\right )^2 T \left (-\frac{n_{\ell_i}}{2} - \frac{n_{\bar{\ell}_i}}{2} - \frac{n_{H_L}}{4} - \frac{n_{H_R}}{4}\right )\nonumber \\
  & - \alpha_R \sum_{ij} \left (y_E^{ij}\right )^2 T \left (-\frac{n_{L_i}}{2} - n_{\bar{E}_j} + \frac{n_{H_R}}{4}\right )  + \alpha_R \sum_{ij} \left (y_N^{ij}\right )^2 T \left (-\frac{n_{L_i}}{2} - n_{\bar{N}_j} - \frac{n_{H_R}}{4}\right )\nonumber \\
  & - \alpha_R y_{\mathcal{E}}^2 T \left ( - \frac{n_{\tilde{L}}}{2} - n_{\mathcal{E}} + \frac{n_{H_R}}{4}\right )  + \alpha_R y_{\mathcal{X}}^2 T \left ( - \frac{n_{\tilde{L}}}{2} - n_{\mathcal{X}} - \frac{n_{H_R}}{4}\right )\nonumber \\
  & + \alpha_R \sum_i \left (g_\mathcal{E}^i\right )^2 T \left (-\frac{n_{\bar{\ell}_i}}{2} - n_\mathcal{E} - \frac{n_{H_R}}{4}\right ),\nonumber\\
  \dot{n}_{\rm PQ} &= \Gamma_{\rm ss}\left (\sum_{j}\left (- n_{q_j} - n_{\bar{q}_j}\right ) -  \dot{\theta}T^2\right ) + \Gamma_{\rm ws} \left (\sum_j \left (-n_{q_j} - n_{\ell_j}\right ) - \frac{c_L}{3} \theta T^2\right ) \nonumber \\
  & + \Gamma_{\rm rs}\left (\sum_j \left (- n_{\bar{q}_j} - n_{\bar{\ell}_j} - n_{L_j}\right ) - n_{\tilde{L}} - \frac{c_{R}}{3} \dot{\theta} T^2\right ). \nonumber
 \end{align}
The indices $i,j$ run from 1 to 3 otherwise explicitly noted. $c_L$ and $c_R$ are the $SU(2)_L$ and $SU(2)_R$ anomaly coefficient of the PQ symmetry normalized by that of the QCD anomaly.
$\Gamma_{\rm ss}$, $\Gamma_{\rm ws}$, and $\Gamma_{\rm rs}$ are QCD, $SU(2)_L$, and $SU(2)_R$ sphaleron rates, respectively.

We obtain different values of $c_B$ depending on if the transfer of the lepton asymmetry of the extra chiral leptons into the SM sector is effective around the $SU(2)_R$ phase transition.
\begin{itemize}
    \item If the portal interactions are not effective around the $SU(2)_R$ phase transition, we may ignore them in these equations. The conserved quantities are the gauge charge $X$, the SM $(B-L)_{\rm SM}$, and a linear combination of lepton numbers defined as $L_{\rm new} \equiv \frac{1}{3} \sum_i (n_{L_i} - n_{\bar{E}_i} - n_{\bar{N}_i}) - (n_{\tilde{L}} - n_{\mathcal{E}} - n_{\mathcal{X}})$. $c_B$ is given by
    \begin{align}
    \label{eq:cB_chiral_1}
        c_B &= -\frac{1}{4} \frac{\sum_i (n_{L_i} - n_{\bar{E}_i} - n_{\bar{N}_i}) - n_{\tilde{L}} + n_{\mathcal{E}} + n_{\mathcal{X}}}{\dot{\theta}T^2} \nonumber \\
        & = -\frac{1}{8} + \frac{1}{12}(c_L + c_R) \simeq 0.125 + 0.083 (c_L + c_R)
    \end{align}
    \item If the portal Yukawa interactions are effective, the only conserved quantity is the gauge charge $X$. $c_B$ is given by
    \begin{align}
    \label{eq:cB_chiral_2}
        c_B &= \frac{1}{4} \frac{\sum_i (\frac{1}{3}(n_{q_i} - n_{\bar{q_i}}) - (n_{\ell_i} - n_{\bar{\ell}_i})+n_{L_i} - n_{\bar{E}_i} - n_{\bar{N}_i}) + n_{\tilde{L}} - n_{\mathcal{E}} - n_{\mathcal{X}}}{\dot{\theta}T^2}\nonumber \\
        & = - \frac{17}{48} + \frac{2}{9}c_L + \frac{1}{4}c_R \simeq -0.354 + 0.222 c_L + 0.250 c_R.
    \end{align}
    For this case, whether or not the transfer by the mass term $m_L^{ij} \bar{\ell}_i L_j$ is effective does not change $c_B$ since it does not change the conservation laws.
    \item If only the transfer by $m_L^{ij} \bar{\ell}_i L_j$ is effective, a conserved quantity other than the gauge charge $X$ is
\begin{align}
    \frac{1}{3}\sum_i (n_{q_i} - n_{\bar{q}_i}) - \sum_i ( n_{\ell_i} - n_{\bar{\ell}_i} + n_{L_i} - n_{\bar{E}_i} - n_{\bar{N}_i}) + 3 (n_{\tilde{L}} - n_{\mathcal{E}} - n_{\mathcal{X}}).
\end{align}
We obtain
\begin{align}
\label{eq:cB_chiral_mass}
    c_B &= \frac{1}{4} \frac{\sum_i (\frac{1}{3}(n_{q_i} - n_{\bar{q_i}}) - (n_{\ell_i} - n_{\bar{\ell}_i})+n_{L_i} - n_{\bar{E}_i} - n_{\bar{N}_i}) + n_{\tilde{L}} - n_{\mathcal{E}} - n_{\mathcal{X}}}{\dot{\theta}T^2}\nonumber \\
    &= -\frac{13}{64} + \frac{1}{8}c_L + \frac{7}{48}c_R \simeq -0.20 + 0.13c_L + 0.15 c_R. 
\end{align}
    \end{itemize}

\subsubsection*{SM Higgs from $\Phi$}
The transport equation is similar to that of the setup with $H_L$ except for the following modifications.
$(n_{H_L} + n_{H_R})/4$ in the SM Yukawa contributions is replaced with $n_\Phi/8$.
Other terms involving $H_L$ is removed.
The equation for $H_L$ is replaced with that for $\Phi$,
\begin{align}
\dot{n}_\Phi &= - \alpha_3 \sum_{ij} \left (y_u^{ij}\right )^2 T \left (-\frac{n_{q_i}}{6} - \frac{n_{\bar{q}_j}}{6} + \frac{n_{\Phi}}{8}\right ) + \alpha_3 \sum_{ij} \left (y_d^{ij}\right )^2 T \left (-\frac{n_{q_i}}{6} - \frac{n_{\bar{q}_j}}{6} - \frac{n_\Phi}{8}\right ) \\
& + \left (\alpha_L + \alpha_R\right ) \sum_{ij} \left (y_l^{ij}\right )^2 T \left (-\frac{n_{\ell_i}}{2} - \frac{n_{\bar{\ell}_j}}{2} - \frac{n_\Phi}{8}\right ) \nonumber \\
&+ \left (\alpha_L + \alpha_R\right ) \sum_{ij} \left (\tilde{y}_l^{ij}\right )^2 T \left (-\frac{n_{\ell_i}}{2} - \frac{n_{\bar{\ell}_j}}{2} + \frac{n_\Phi}{8}\right ). \nonumber
\end{align}
$\Phi$ is a complex field, so we may also include the contributions from $\Phi$-number violating potential terms such as $\Phi^2$, but
that does not violate symmetry additionally and adding these terms does not change the result.

The result depends on whether or not the portal interaction is effective around the $SU(2)_R$ phase transition.
    \begin{itemize}
    \item If the portal interactions are not effective around the $SU(2)_R$ phase transition, $c_B$ is the same as Eq.~(\ref{eq:cB_chiral_1}). 
    \item If the portal interactions are effective except for the $m_L^{ij} \bar{\ell}_i L_j$ term, the following quantity is conserved,
    \begin{align}
        n_{\tilde{L}} - n_{\mathcal{E}} - n_\mathcal{X} - \sum_i ( n_{\ell_i} - n_{\bar{\ell}_i}) + \frac{1}{3}\sum_i (n_{q_i} - n_{\bar{q}_i}) - \frac{1}{3} \sum_i (n_{L_i} - n_{\bar{E}_i} - n_{\bar{N}_i}).
    \end{align}
    The resultant $c_B$ is
    \begin{align}
        c_B &= \frac{1}{4} \frac{\sum_i (\frac{1}{3}(n_{q_i} - n_{\bar{q_i}}) - (n_{\ell_i} - n_{\bar{\ell}_i})+n_{L_i} - n_{\bar{E}_i} - n_{\bar{N}_i}) + n_{\tilde{L}} - n_{\mathcal{E}} - n_{\mathcal{X}}}{\dot{\theta}T^2}\nonumber \\
        &=- \frac{549}{4040}  + \frac{137}{1515}c_L + \frac{55}{606}c_R \simeq -0.136 + 0.090 c_L + 0.091 c_R.
    \end{align}
    \item If only the transfer by $m_L^{ij} \bar{\ell}_i L_j$ is effective, $c_B$ is the same as Eq.~(\ref{eq:cB_chiral_mass}).
    \item If all of the portal interactions are effective, the only conserved quantity is the gauge charge $X$. The absence of other conservation laws makes $c_B$ much larger,
    \begin{align}
        c_B &= \frac{1}{4} \frac{\sum_i (\frac{1}{3}(n_{q_i} - n_{\bar{q_i}}) - (n_{\ell_i} - n_{\bar{\ell}_i})+n_{L_i} - n_{\bar{E}_i} - n_{\bar{N}_i}) + n_{\tilde{L}} - n_{\mathcal{E}} - n_{\mathcal{X}}}{\dot{\theta}T^2}\nonumber \\
        &= -\frac{49}{152} + \frac{23}{114}c_L + \frac{13}{57}c_R \simeq -0.322 + 0.202 c_L + 0.228 c_R.
    \end{align}
\end{itemize}

\subsection{Singlet right-handed tau}

The transport equation for the model with an $SU(2)_R$ singlet right-handed tau in Sec.~\ref{subsec:extra singlet} is given by
\begin{align}
  \label{eq:boltzmann eq}
  \dot{n}_{q_{i}} =
  &\ \alpha_3\sum_{j}\left (y_{u}^{ij}\right )^{2}T\left (-\frac{n_{q_{i}}}{6} -  \frac{n_{\bar{q}_{j}}}{6} + \frac{n_{H_{L}}}{4} + \frac{n_{H_R}}{4}\right )  \\
  &+ \alpha_3\sum_{j}\left (y_{d}^{ij}\right )^{2}T\left (-\frac{n_{q_{i}}}{6} -  \frac{n_{\bar{q}_{j}}}{6} - \frac{n_{H_{L}}}{4} - \frac{n_{H_R}}{4}\right )\nonumber \\
   & + 2 \Gamma_{\rm ss}\left (\sum_{j}\left (- n_{q_j} - n_{\bar{q}_j}\right ) - \frac{1}{2}\dot{\theta}T^2\right ) + 3 \Gamma_{\rm ws}\left (\sum_{j}\left (-n_{q_j} - n_{\ell_j}\right ) - \frac{c_L}{3}\dot{\theta}T^{2}\right ), \nonumber \\
  \dot{n}_{\bar{q}_{i}} =
  &\ \alpha_3\sum_{j}\left (y_u^{ji}\right )^2T\left (-\frac{n_{q_{j}}}{6} - \frac{n_{\bar{q}_i}}{6} + \frac{n_{H_L}}{4} + \frac{n_{H_R}}{4}\right ) \nonumber\\
  &+ \alpha_3\sum_{j}\left (y_d^{ji}\right )^2T\left (-\frac{n_{q_{j}}}{6} - \frac{n_{\bar{q}_{i}}}{6} - \frac{n_{H_L}}{4} - \frac{n_{H_R}}{4}\right )\nonumber \\
  & + 2 \Gamma_{\rm ss}\left (\sum_{j}\left (- n_{q_j} - n_{\bar{q}_j}\right )  - \frac{1}{2}\dot{\theta}T^2\right ) + 3 \Gamma_{\rm rs}\left (\sum_{j}\left (-n_{\bar{q}_{j}} - n_{\bar{\ell}_{j}}\right ) - \frac{c_{R}}{3}\dot{\theta}T^{2}\right ), \nonumber \\
  \dot{n}_{\ell_{i}} =
  &\ \left (\alpha_L + \alpha_R\right )\left (y_e^{ii}\right )^2T\left (-\frac{n_{\ell_i}}{2} - \frac{n_{\bar{\ell}_i}}{2} - \frac{n_{H_L}}{4} - \frac{n_{H_R}}{4}\right ) +\alpha_L \left (y_{e}^{i3}\right )^2T\left (-\frac{n_{\ell_i}}{2} - n_{\bar{\tau}} - \frac{n_{H_L}}{4}\right )\nonumber  \\
  & + \Gamma_{\rm ws}\left (\sum_{j}\left (-n_{q_j} - n_{\ell_j}\right ) - \frac{c_L}{3}\dot{\theta}T^{2}\right ),\nonumber \\
  \dot{n}_{\bar{\ell}_{i=1,2}} =
  &\ \left (\alpha_L + \alpha_R\right )\left (y_e^{ii}\right )^{2}T\left (-\frac{n_{\ell_{i}}}{2} - \frac{n_{\bar{\ell}_{i}}}{2} - \frac{n_{H_L}}{4} - \frac{n_{H_R}}{4}\right ) + \Gamma_{\rm rs}\left (\sum_{j}\left (-n_{\bar{q}_{j}} - n_{\bar{\ell}_{j}}\right ) - \frac{c_{R}}{3}\dot{\theta}T^{2}\right ), \nonumber \\
  \dot{n}_{\bar{\ell}_{3}} =
  &\ \alpha_R y_{\tau^{\prime}}^2T\left (-\frac{n_{\bar{\ell}_3}}{2} - n_{\tau^{\prime}} - \frac{n_{H_R}}{4}\right ) + \Gamma_{\rm rs}\left (\sum_{j}\left (-n_{\bar{q}_{j}} - n_{\bar{\ell}_{j}}\right ) - \frac{c_{R}}{3}\dot{\theta}T^{2}\right ), \nonumber \\
  \dot{n}_{\bar{\tau}} =
  &\ \alpha_L  \sum_i \left (y_{e}^{i3}\right )^2T\left (-\frac{n_{\ell_i}}{2} - n_{\bar{\tau}} - \frac{n_{H_L}}{4}\right ), \nonumber  \\
  \dot{n}_{\tau^{\prime}} =
  & \alpha_R \left (y_{\tau^{\prime}}\right )^2T\left (- \frac{n_{\bar{\ell}_3}}{2} - n_{\tau^{\prime}} - \frac{n_{H_R}}{4}\right ),\nonumber \\
  \dot{n}_{H_L} =
  &-\alpha_3\sum_{i,j}\left (y_{u}^{ij}\right )^{2}T\left (-\frac{n_{q_{i}}}{6} -  \frac{n_{\bar{q}_{j}}}{6} + \frac{n_{H_{L}}}{4} + \frac{n_{H_R}}{4}\right ) \nonumber \\
  & + \alpha_3\sum_{i,j}\left (y_{d}^{ij}\right )^{2}T\left (-\frac{n_{q_{i}}}{6} -  \frac{n_{\bar{q}_{j}}}{6} - \frac{n_{H_{L}}}{4} - \frac{n_{H_R}}{4}\right )\nonumber \\
   & + \left (\alpha_L + \alpha_R\right )\sum_{i=1,2}\left (y_e^{ii}\right )^2T\left (-\frac{n_{\ell_i}}{2} - \frac{n_{\bar{\ell}_i}}{2} - \frac{n_{H_L}}{4} - \frac{n_{H_R}}{4}\right ) \nonumber \\
   &+\alpha_L \sum_i \left (y_{e}^{i3}\right )^2T\left (-\frac{n_{\ell_i}}{2} - n_{\bar{\tau}} - \frac{n_{H_L}}{4}\right ),\nonumber  \\
  \dot{n}_{H_R} =
  & - \alpha_3\sum_{i,j}\left (y_u^{ij}\right )^2T\left (-\frac{n_{q_i}}{6} - \frac{n_{\bar{q}_j}}{6} + \frac{n_{H_L}}{4} + \frac{n_{H_R}}{4}\right ) \nonumber \\
  &+ \alpha_3\sum_{i,j}\left (y_d^{ij}\right )^2T\left (-\frac{n_{q_i}}{6} - \frac{n_{\bar{q}_j}}{6} - \frac{n_{H_L}}{4} - \frac{n_{H_R}}{4}\right )\nonumber \\
  & + \left (\alpha_L + \alpha_R\right )\sum_{i=1,2}\left (y_e^{ii}\right )^2T\left (-\frac{n_{\ell_i}}{2} - \frac{n_{\bar{\ell}_i}}{2} - \frac{n_{H_L}}{4} - \frac{n_{H_R}}{4}\right ) \nonumber \\
  & + \alpha_R \left (y_{\tau^{\prime}}\right )^2 T \left (-\frac{n_{\bar{\ell}_3}}{2} - n_{\tau^{\prime}} - \frac{n_{H_R}}{4}\right ),\nonumber  \\
  \dot{n}_{P} =
  & \Gamma_{\rm ss}\left (\sum_{i}\left (- n_{q_i} - n_{\bar{q}_i}\right ) - \frac{1}{2}\dot{\theta}T^2\right )\nonumber \\
  & + c_L\Gamma_{\rm ws}\left (\sum_{j}\left (-n_{q_j} - n_{\ell_j}\right ) - \frac{c_L}{3}\dot{\theta}T^{2}\right )
   + c_{R}\Gamma_{\rm rs}\left (\sum_{j}\left (-n_{\bar{q}_{j}} - n_{\bar{\ell}_{j}}\right ) - \frac{c_{R}}{3}\dot{\theta}T^{2}\right ). \nonumber 
\end{align}

The conserved quantities are the gauge charge $X$ and the total $B-L$ number, and we obtain
\begin{align}
    c_B &= -\frac{1}{4}\frac{(n_{\bar{\ell}_3} - n_{\tau^\prime})}{\dot{\theta}T^2}\nonumber \\
    &= \frac{155}{1056} - \frac{37}{396}c_W - \frac{9}{88}c_R \simeq 0.147 - 0.093 c_L -0.102 c_R.
\end{align}

In the above estimation, we assume that $\bar{\tau}$ couples to all of $\ell_i$ via the Yukawa couplings $y_e^{i3}$. 
If $\bar{\tau}$ only couples to $\ell_3$, there are two extra conserved quantities: $L_{12} \equiv  (n_{\ell_1} - n_{\bar{\ell}_1}) - (n_{\ell_2} - n_{\bar{\ell}_2})$, $L_{13} \equiv (n_{\ell_1} - n_{\bar{\ell}_1}) - (n_{\ell_3} - n_{\bar{\ell}_3} + n_{\tau^{\prime}} - n_{\bar{\tau}})$
. Imposing these conserved quantities, we obtain smaller $c_B$,
\begin{align}
  \label{eq:cB for extra singlet model}
  c_B &= -\frac{1}{4}\frac{(n_{\bar{\ell}_3} - n_{\tau^{\prime}})}{\dot{\theta}T^2}\nonumber \\
  &= \frac{3}{32} - \frac{29}{492}c_L - \frac{65}{984}c_R \simeq 0.094 - 0.059 c_L - 0.066 c_R.
\end{align}

\subsection{Extra vector-like leptons}

The transport equations are
\begin{align}
  \label{eq:vectorlike transport}
    \dot{n}_{q_{i}} &= \sum_{j} \alpha_3 \left (y_{q}^{ij}\right )^{2}T \left (-\frac{n_{q_i}}{6} - \frac{n_{\bar{q}_j}}{6} - \frac{n_\Phi}{8}\right ) + \sum_{j} \alpha_3 \left (\tilde{y}_{q}^{ij}\right )^{2}T \left (-\frac{n_{q_i}}{6} - \frac{n_{\bar{q}_j}}{6} + \frac{n_\Phi}{8}\right )\nonumber \\
    & + 2 \Gamma_{\rm ss}\left (\sum_{j}\left (-n_{q_{j}} - n_{\bar{q}_{j}}\right ) - c_{g}\dot{\theta} T^2\right )  + 3 \Gamma_{\rm ws}\left (\sum_{j}\left (-n_{q_{j}} - n_{l_{j}}\right ) - \frac{c_L}{3} \dot{\theta} T^{2}\right ), \nonumber \\
    \dot{n}_{\bar{q}_{i}} & = \sum_{j} \alpha_3 \left (y_{q}^{ij}\right )^{2}T \left (-\frac{n_{q_i}}{6} - \frac{n_{\bar{q}_j}}{6} - \frac{n_\Phi}{8}\right ) + \sum_{j} \alpha_3 \left (\tilde{y}_{q}^{ij}\right )^{2}T \left (-\frac{n_{q_i}}{6} - \frac{n_{\bar{q}_j}}{6} + \frac{n_\Phi}{8}\right )\nonumber \\
   &  + 2 \Gamma_{\rm ss}\left (\sum_{j}\left (-n_{q_{j}} - n_{\bar{q}_{j}}\right ) - c_{g}\dot{\theta} T^2\right )+ 3 \Gamma_{\rm rs}\left (\sum_{j}\left (-n_{\bar{q}_{j}} - n_{\bar{\ell}_{j}}\right ) - n_{L} - n_{\bar{L}} - \frac{c_{R}}{3} \dot{\theta} T^{2}\right ), \nonumber \\
    \dot{n}_{\ell_{i}} &= \left (\alpha_{R} + \alpha_{L}\right )\sum_{j}\left (y_{\ell}^{ij}\right )^{2}\left (-\frac{n_{\ell_i}}{2} - \frac{n_{\bar{\ell}_j}}{2} - \frac{n_\Phi}{8}\right ) +  \left (\alpha_{R} + \alpha_{L}\right )\sum_{j}\left (\tilde{y}_{\ell}^{ij}\right )^{2}\left (-\frac{n_{\ell_i}}{2} - \frac{n_{\bar{\ell}_j}}{2} + \frac{n_\Phi}{8}\right )\nonumber \\
     & + \Gamma_{\rm ws}\left (\sum_{j}\left (-n_{q_{j}} - n_{l_{j}}\right ) - \frac{c_L}{3} \dot{\theta} T^{2}\right )\nonumber \\
     & + \left (\alpha_L + \alpha_R\right ) \left (g_L^i\right )^2 T \left (-\frac{n_{\ell_i}}{2} - \frac{n_{\bar{L}}}{2} - \frac{n_\Phi}{8}\right ), \nonumber \\
    \dot{n}_{\bar{\ell}_{i}}  &= \left (\alpha_{R} + \alpha_{L}\right )\sum_{j}\left (y_{\ell}^{ij}\right )^{2}\left (-\frac{n_{\ell_j}}{2} - \frac{n_{\bar{\ell}_i}}{2} - \frac{n_\Phi}{8}\right ) +  \left (\alpha_{R} + \alpha_{L}\right )\sum_{j}\left (\tilde{y}_{\ell}^{ij}\right )^{2}\left (-\frac{n_{\ell_j}}{2} - \frac{n_{\bar{\ell}_i}}{2} + \frac{n_\Phi}{8}\right )\nonumber \\
      & + \Gamma_{\rm rs}\left (\sum_{j}\left (-n_{\bar{q}_j} - n_{\bar{\ell}_j}\right ) -n_{L} - n_{\bar{L}} - \frac{c_{R}}{3} \dot{\theta} T^{2}\right )\nonumber \\
      & + \alpha_R \left (g_E^i\right )^2 T \left (-\frac{n_{\bar{\ell}_i}}{2} - n_E - \frac{n_{H_R}}{4}\right ) + \alpha_R \frac{\left (m_L^i\right )^2}{T}\left (-\frac{n_{\bar{\ell}_i}}{2} - \frac{n_L}{2}\right ),\nonumber \\
    \dot{n}_{L} &= \alpha_{R}y_L^2T\left (-\frac{n_{L}}{2} - n_{\bar{E}} + \frac{n_{H_R}}{4}\right )\nonumber \\
     & + \Gamma_{\rm rs}\left (\sum_{j}\left (-n_{\bar{q}_{j}} - n_{\bar{\ell}_{j}}\right ) -n_{L} - n_{\bar{L}} - \frac{c_{R}}{3} \dot{\theta} T^{2}\right )\nonumber \\
     & + \sum_i \alpha_R \frac{\left (m_L^i\right )^2}{T}\left (-\frac{n_{\bar{\ell}_i}}{2} - \frac{n_L}{2}\right ),\nonumber \\
    \dot{n}_{\bar{E}} & = \alpha_{R}y_L^2T\left (-\frac{n_{L}}{2} - n_{\bar{E}} + \frac{n_{H_R}}{4}\right ),\nonumber \\
    \dot{n}_{\bar{L}} &= \alpha_{R}y_{\bar{L}}^2T\left (-\frac{n_{\bar{L}}}{2} - n_{E} - \frac{n_{H_R}}{4}\right )\nonumber \\
     & + \Gamma_{\rm rs}\left (\sum_{j}\left (-n_{\bar{q}_{j}} - n_{\bar{\ell}_{j}}\right )  -n_{L} - n_{\bar{L}} - \frac{c_{R}}{3} \dot{\theta} T^{2}\right )\nonumber \\
      & + \left (\alpha_L + \alpha_R\right ) \sum_i \left (g_L^i\right )^2 T \left (-\frac{n_{\ell_i}}{2} - \frac{n_{\bar{L}}}{2} - \frac{n_\Phi}{8}\right ),\nonumber \\
    \dot{n}_{E} &= \alpha_{R}y_{\bar{L}}^2T\left (-\frac{n_{\bar{L}}}{2} - n_{E} - \frac{n_{H_R}}{4}\right )\nonumber \\
    & + \alpha_R \sum_i \left (g_E^i\right )^2 T \left (-\frac{n_{\bar{\ell}_i}}{2} - n_E - \frac{n_{H_R}}{4}\right ),\nonumber \\
    \dot{n}_{\Phi} &= \sum_{i,j}\alpha_3 \left (y_{q}^{ij}\right )^{2}T \left (-\frac{n_{q_i}}{6} - \frac{n_{\bar{q}_j}}{6} - \frac{n_\Phi}{8}\right )
    - \sum_{ij} \alpha_3 \left (\tilde{y}_{q}^{ij}\right )^{2}T \left (-\frac{n_{q_i}}{6} - \frac{n_{\bar{q}_j}}{6} + \frac{n_\Phi}{8}\right )\nonumber \\
     & +  \left (\alpha_{R} + \alpha_{L}\right )\sum_{i,j}\left (y_{\ell}^{ij}\right )^{2}\left (-\frac{n_{\ell_j}}{2} - \frac{n_{\bar{\ell}_j}}{2} - \frac{n_\Phi}{8}\right )
     - \left (\alpha_{R} + \alpha_{L}\right )\sum_{ij}\left (\tilde{y}_{\ell}^{ij}\right )^{2}\left (-\frac{n_{\ell_i}}{2} - \frac{n_{\bar{\ell}_j}}{2} + \frac{n_\Phi}{8}\right ) \nonumber \\
                          & + \left (\alpha_L + \alpha_R\right ) \sum_{ij} \left (g_L^i\right )^2 T \left (-\frac{n_{\ell_i}}{2} - \frac{n_{\bar{L}}}{2} - \frac{n_\Phi}{8}\right ),\nonumber \\
    \dot{n}_{H_{R}} &= - \alpha_{R}y_L^2T\left (-\frac{n_{L}}{2} - n_{\bar{E}} + \frac{n_{H_R}}{4}\right ) + \alpha_{R}y_{\bar{L}}^2T\left (-\frac{n_{\bar{L}}}{2} - n_{E} - \frac{n_{H_R}}{4}\right )\nonumber\\
                          & + \alpha_R \sum_i \left (g_E^i\right )^2 T \left (-\frac{n_{\bar{\ell}_i}}{2} - n_E - \frac{n_{H_R}}{4}\right ),\nonumber  \\
    \dot{n}_{\rm PQ} &= \Gamma_{\rm ss}\left (\sum_{j}\left (-n_{q_{j}} - n_{\bar{q}_{j}}\right ) - c_{g}\dot{\theta} T^2\right ) +\Gamma_{\rm ws}\left (\sum_{j}\left (-n_{q_{j}} - n_{\ell_j}\right ) - \frac{c_L}{3} \dot{\theta} T^{2}\right )\nonumber \\
    &+\Gamma_{\rm rs}\left (\sum_{j}\left (-n_{\bar{q}_j} - n_{\bar{\ell}_j}\right )  -n_{L} - n_{\bar{L}} - \frac{c_{R}}{3} \dot{\theta} T^{2}\right ). \nonumber 
  \end{align}

When the portal terms $m_L^i L \bar{\ell}_i$, $g_L^i \bar{L} \Phi \ell_i$ and $g_E^i \bar{\ell}_i H_R E$ are not effective around the $SU(2)_R$ phase transition,
the conserved quantities are the $X$ charge, the SM $B-L$, $L_{\rm 12}$, $L_{\rm 13}$ defined in the previous subsection, plus $L_{\rm exotic} \equiv (n_L - n_{\bar{E}}) - (n_{\bar{L}} - n_E)$. We obtain
\begin{align}
    c_B &= \pm \frac{1}{4}\frac{n_{\bar{L}} - n_{\bar{E}}}{\dot{\theta}T^2} = \mp \frac{1}{4}\frac{n_{\bar{L}} - n_E}{\dot{\theta}T^2}\nonumber \\
    &= \pm \left(\frac{3}{32} - \frac{1}{16}(c_L + c_R)\right) \simeq \pm \left(0.094 - 0.063\left(c_L + c_R\right)\right).
\end{align}

If the transfer of the asymmetry of $(\bar{L},E)$ into the SM sector is effective around the $SU(2)_R$ phase transition, the only conserved quantities are the $X$ charge and the total $B-L$ number, and we obtain
\begin{align}
    c_B &= \frac{1}{4} \frac{n_L - n_{\bar{E}}}{\dot{\theta}T^2}\nonumber \\
    &= \frac{31}{192} - \frac{1}{9}c_L - \frac{5}{48}c_R \simeq 0.161 - 0.111 c_L - 0.104 c_R.
\end{align}
If the transfer of the asymmetry of $(L, \bar{E})$ into the SM sector is effective around the $SU(2)_R$ phase transition,
the conserved quantity is still the $X$ charge and total $B-L$. We obtain
\begin{align}
    c_B &= \frac{1}{4} \frac{n_{\bar{L}} - n_{E}}{\dot{\theta}T^2}\nonumber \\
    &= \frac{29}{192} - \frac{7}{72}c_L - \frac{5}{48}c_R \simeq 0.151 - 0.097 c_L - 0.104 c_R.
\end{align}

\section{Details of effective potential}\label{sec:effective potential}

This appendix shows the details of the decoupling of the $SU(2)_R$ sphaleron process during the $SU(2)_R$ phase transition.
To obtain the sphaleron decoupling temperature as a function of the model parameters, we compute the effective potential of $H_R=(0, \phi)^{t}$.

The general form of total effective potential is
\begin{align}
  \label{eq:total effective potential}
  V_{\rm eff}(\phi,T) = V_0(\phi) + V_{\rm CW}(\phi) + V_{\rm FT}(\phi, T).
\end{align}
The tree-level Higgs potential is
\begin{align}
  \label{eq:tree level effective potential}
  V_0 = - 2\lambda_R v_R^2 \phi^2 + \lambda_R\phi^{4},
\end{align}
where $v_R$ is the tree-level vev.
We compute the correction to the  potential up to one-loop level, including zero-temperature Coleman-Weinberg corrections and finite-temperature corrections~\cite{Quiros:1999jp}.

We adopt the renormalization condition that keeps $v_R$ and the Higgs mass around the minimum of the potential unchanged by quantum corrections at zero temperature, namely, $V^{\prime}(\phi = v_R)|_{T=0} = 0$ and $V^{\prime \prime}(\phi = v_R)|_{T=0} = 8 \lambda_R v_R^2$ \cite{Dine:1992wr}. Under this renormalization condition, the 1-loop Coleman-Weinberg correction at zero temperature is
\begin{align}
  \label{eq:1-loop formula}
  V_{\rm CW}(\phi) = 2B v_R^2 \phi^2 - \frac{3}{2} B \phi^4 + B \phi^4 \log(\frac{\phi^2}{v_R^2}).
\end{align}
The zero-temperature correction from $\lambda_R$ is always suppressed by a loop factor in comparison with the tree-level potential and hence is negligible.
We only include gauge boson and fermion contributions, for which the coefficient $B$ is given by
\begin{align}
  \label{eq:B definition}
  B = & \frac{1}{64 \pi^2 v_R^4}\left(\sum_{i={\rm bosons}}n_i m_i^4 - \sum_{i = {\rm fermions}} n_i m_i^4\right) \nonumber \\
   = & \frac{1}{64 \pi^2 v_R^4}\left(6 m_{W_R}^4(v_R) + 3 m_{Z^{\prime}}^4(v_R) - \sum_{f} n_f m_f^4(v_R) \right).
\end{align}
Here $n_i$ are the degree of freedom and $m_i$ are the masses of the corresponding particles at $\phi=v_R$. Note that the coefficient $B$ is independent of the field value.

Next, we calculate the finite-temperature corrections. Because of the large occupation number at zero energy, bosonic contributions need resummation. Here we employ the Parwani resummation method~\cite{Parwani:1991gq} and add thermal self-energy in the high-temperature expansion to
the masses of particles inside the loop. The finite-temperature correction at one-loop level is then given by
\begin{align}
  \label{eq:finite T effective potential formula}
  V_{\rm FT}(\phi, T) = \frac{T^4}{2\pi}\left (\sum_{i=bosons} n_i J_B\left (\frac{m_i(\phi, T)^2}{T^2}\right ) - \sum_{i=fermions}n_i J_F\left (\frac{m_i(\phi)^2}{T^2}\right ) \right),
\end{align}
where $m_i(\phi)$ are the field-dependent masses at $T=0$ and $m_i(\phi, T)$ are masses at finite temperature $T$. Summation over bosons contains both gauge bosons and scalar bosons. (Notice that unlike the $T=0$ case, the finite temperature correction from the scalars are not necessarily negligible). The thermal functions $J_B$ and $J_F$ are defined as
\begin{align}
  \label{eq:JB and JF}
  J_{B/F}(r^2) = \int_0^{\infty} \rmd x x^2 \log (1 \pm e^{\sqrt{x^2 + r^2}}).
\end{align}

The thermal potential is then given by
\begin{align}
  \label{eq:B coefficient}
    V_{\rm FT}(\phi,T) = \frac{T^4}{2\pi}&\left (4 J_B\left (\frac{m_{W_{R,\bot}}^{2}(\phi,T)}{T^2}\right ) + 2J_B \left(\frac{m_{W_{R,\parallel}}^2(\phi,T)}{T^2} \right)\right. \nonumber \\
    &\left. + 2 J_B\left (\frac{m_{Z^{\prime}_\bot}^2(\phi,T)}{T^{2}}\right ) + J_B \left ( \frac{m_{Z^\prime_\parallel}^2(\phi,T)}{T^2}\right ) + \right. \nonumber \\ 
    &\left. + J_B \left (\frac{m_{h_R}^2(\phi,T)}{T^2}\right) + 3 J_B \left( \frac{m_\chi^2(\phi,T)}{T^2}\right)\right. \nonumber \\
   &-\left.  \sum_f n_f J_F\left ( \frac{m_f^2(\phi)}{T^2}\right ) \right ),
\end{align}
where $\bot$ and $\parallel$ denote transverse and longitudinal components, respectively.
The masses of physical $h_R$, Nambu-Goldstone bosons $\chi$, and the charged $W_R$ boson are given by~\cite{Carrington:1991hz,Weldon:1989ys}
\begin{align}
  \label{eq:Higgs mass}
  m_{h_R}^2(\phi,T) =& 2 \lambda_R(3 \phi^2 -  v_R^2) + \frac{1}{2}\lambda_RT^2 + \frac{1}{2}y_{t}^2 T^2 + \frac{1}{8}g_R^2T^2 + \frac{1}{16}(g_R^2 + g_X^2)T^2, \\
  \label{eq:gs boson mass}
  m_\chi^2(\phi,T) =& 2\lambda_R( \phi^2 - v_R^2) + \frac{1}{2}\lambda_RT^2 + \frac{1}{2}y_{t}^2 T^2 + \frac{1}{8}g_R^2T^2 + \frac{1}{16}(g_R^2 + g_X^2)T^2, \\
  \label{eq:W12 mass}
  m_{W_{R}}^2(\phi,T) =& \frac{1}{2} g_R^2 \phi^2 + \Pi_{W},
\end{align}
where $\Pi_W$ is the thermal self-energy shown later. The neutral gauge boson mass matrix of $W_R^3$ and $B_X$ is
\begin{align}
  \label{eq:neutral gauge boson mass}
  \mathcal{M}_{\rm neutral} =
  \begin{pmatrix}
    \frac{1}{2}g_R^2 \phi^2 + \Pi_W & - \frac{1}{2}g_R g_X \phi^2 \\
    - \frac{1}{2} g_R g_X \phi^2 & \frac{1}{2} g_X^2 \phi^2 + \Pi_{X}
  \end{pmatrix},
\end{align}
where $\Pi_X$ is the thermal self-energy for the $U(1)_X$ gauge boson $B_X$. The mass eigenstates are nothing but the $Z^{\prime}$ and SM $B$ gauge bosons with masses $m_{Z^{\prime}}(\phi,T)$ and $m_B(\phi,T)$. At zero temperature, $m_{Z^{\prime}}^2(\phi) = (g_R^2 + g_X^2)\phi^2/2$ and $m_B = 0$.

For the purpose of resummation, we are interested in the thermal self energy at $\omega\rightarrow0$. In that limit, among the three components of a gauge boson, only the longitudinal mode receives thermal corrections at the one-loop order~\cite{Linde:1980ts,Gross:1980br}. The thermal corrections for the model with extra chiral leptons are~\cite{Carrington:1991hz,Weldon:1989ys}
\begin{align}
  \label{eq:gauge thermal mass with chiral}
  \Pi_{W,\parallel} = \frac{13}{6}g_R^2 T^2, \Pi_{X,\parallel} = \frac{39}{12} g_X^2 T^2.
\end{align}
For the model with a singlet right-handed tau,
\begin{align}
  \label{eq:gauge thermal mass for one extra singlet}
  \Pi_{W,\parallel} = \frac{11}{6}g_R^2 T^2, \Pi_{X,\parallel} = \frac{29}{18}g_X^2 T^2.
\end{align}
For the model with extra vector-like leptons,
\begin{align}
  \label{eq:gauge thermal mass with vectorlike}
  \Pi_{W,\parallel} = 2 g_R^2 T^2, \Pi_{X,\parallel} = \frac{5}{4} g_X^2 T^2.
\end{align}
This small model-dependence of the thermal self energy, however, does not change the prediction on $m_{W_R}$ more than 1\%.

\bibliographystyle{JHEP}
\bibliography{Axiogenesis_SU2R}

\end{document}